\documentclass[10pt,journal,compsoc]{IEEEtran}
\usepackage{cite}
\usepackage{amsmath,amssymb,amsfonts}
\usepackage{algorithmic}
\usepackage{graphicx}
\usepackage{textcomp}
\usepackage{xcolor}
\usepackage{import}
\usepackage{todonotes}
\usepackage{url}
\usepackage{acronym}
\usepackage{amsthm}
\newtheorem{definition}{Definition}
\newtheorem{theorem}{Theorem}
\usepackage[linesnumbered,ruled]{algorithm2e}
\usepackage{array}
\usepackage{silence}
\usepackage{pifont}
\usepackage{wasysym} 
\def\BibTeX{{\rm B\kern-.05em{\sc i\kern-.025em b}\kern-.08em
    T\kern-.1667em\lower.7ex\hbox{E}\kern-.125emX}}

\acrodef{RID}{Remote ID}
\acrodef{FAA}{Federal Aviation Administration}
\acrodef{UAV}{Unmanned Aerial Vehicle}
\acrodef{RFC}{Request for Comment}
\acrodef{WG}{Working Group}
\acrodef{PKE}{Public Key Encryption}
\acrodef{CSP}{Cloud Service Provider}
\acrodef{CI}{Critical Infrastructure}
\acrodef{NFZ}{No-Fly Zone}
\acrodef{TTP}{Trusted Third Party}
\acrodef{CS}{Control Station}
\acrodef{DRIP}{Drone Remote ID Protocol}
\acrodef{EC}{Elliptic Curve}
\acrodef{ECIES}{Elliptic Curve Integrated Encryption Scheme}
\acrodef{SP}{Service Provider}
\acrodef{EV}{Electric Vehicle}
\acrodef{SIM}{Subscriber Identity Module}
\acrodef{DaaS}{Drone-as-a-Service}
\acrodef{RF}{Radio Frequency}
\acrodef{PbIR}{Public Information Registry}
\acrodef{PvIR}{Private Information Registry}
\acrodef{LBS}{Location-Based Service}
\acrodef{EASA}{European Union Aviation Safety Agency}

\acrodef{KAF}{Key Agreement Function}
\acrodef{KDF}{Key Derivation Function}
\acrodef{MAC}{Message Authentication Code}
\acrodef{PIM}{Planar Isotropic Mechanism}
\acrodef{WA}{Warning Area}
\acrodef{NIST}{National Institute of Standards and Technology}
\acrodef{OS}{Operating System}
\acrodef{RPI}{Raspberry PI 3 Model B+}
\acrodef{UID}{Unique Identifier}
\acrodef{RAM}{Random Access Memory}
\acrodef{GNSS}{Global Navigation Satellite System}
\acrodef{GPS}{Global Positioning System}
\acrodef{PC}{Personal Computer}
\acrodef{VM}{Virtual Machine}
\acrodef{MTU}{Maximum Transmission Unit}
\acrodef{TN}{True Negative}
\acrodef{FP}{False Positive}
\acrodef{FN}{False Negative}
\acrodef{TP}{True Positive}
\acrodef{PoC}{Proof-of-Concept}
\acrodef{RQ}{Research Question}
\acrodef{USS}{Unmanned Service Supplier}
\acrodef{MS}{Monitoring Station}
\acrodef{RPAS}{Remotely-Piloted Aircraft System}

\newcommand{\sol}{OLO-RID}

\newcolumntype{P}[1]{>{\centering\arraybackslash}p{#1}}

\begin{document}


\title{Obfuscated Location Disclosure for\\ Remote ID Enabled Drones}

\author{Alessandro Brighente, Mauro Conti, Matthijs Schotsman, Savio Sciancalepore
\IEEEcompsocitemizethanks{
\IEEEcompsocthanksitem Alessandro Brighente and Mauro Conti are with the University of Padova, Department of Information Engineering, Italy.
e-mails: \{alessandro.brighente, mauro.conti\}@unipd.it.
\IEEEcompsocthanksitem Matthijs Schotsman and Savio Sciancalepore are with the Eindhoven University of Technology (TU/e), Department of Mathematics and Computer Science, Eindhoven, The Netherlands.
e-mails: m.schotsman@student.tue.nl, s.sciancalepore@tue.nl.\\
Alphabetical order of the authors.
}
}

\maketitle
\begin{abstract}
\footnote{This is a personal copy of the authors. Not for redistribution. The final version of the paper will be available soon through the digital library IEEExplore at \url{https://ieeexplore.ieee.org/document/11230512/}.}
The Remote ID (RID) regulation recently introduced by several aviation authorities worldwide (including the US and EU) forces commercial drones to regularly (max. every second) broadcast plain-text messages on the wireless channel, providing information about the drone identifier and current location, among others. Although these regulations increase the accountability of drone operations and improve traffic management, they allow malicious users to track drones via the disclosed information, possibly leading to drone capture and severe privacy leaks. \\
In this paper, we propose Obfuscated Location disclOsure for RID-enabled drones (OLO-RID), a solution modifying and extending the RID regulation while preserving drones' location privacy. Rather than disclosing the actual drone's location, drones equipped with OLO-RID disclose a differentially private obfuscated location. OLO-RID also extends RID messages with encrypted location information, accessible only by authorized entities and valuable to obtain the current drone's location in safety-critical use cases. We design, implement, and deploy OLO-RID on a Raspberry Pi 3 and release the code of our implementation as open-source. We also perform an extensive performance assessment of the runtime overhead of our solution in terms of processing, communication, memory, and energy consumption. We show that OLO-RID can generate RID messages on a constrained device in less than $0.16$~s while also requiring a minimal energy toll on a relevant device ($0.0236\%$ of energy for a DJI Mini 2). We also evaluate the utility of the proposed approach in the context of three reference use cases involving the drones' location usage, demonstrating minimal performance degradation when trading off location privacy and utility for next-generation RID-compliant drone ecosystems.

\end{abstract}
\section{Introduction}
\label{sec:introduction}

The adoption of \acp{UAV}, a.k.a. drones\footnote{In this paper, the terms Unmanned Aerial Vehicles (UAVs) and drones are used interchangeably.}, is increasing today in many application areas, e.g., search-and-rescue, surveillance, goods delivery, and agriculture~\cite{Shakhatreh2019_access}. Such pervasive diffusion is also supported by reports of leading market companies, forecasting over 9.6 million consumer drone unit shipments globally by 2030, with 865 million devices already deployed in 2023~\cite{statista_dronesMarket},~\cite{statista_dronesUS}.

At the same time, relevant regional aviation authorities need to integrate drones into the local airspaces to enhance monitoring and accountability of drone operations. To this aim, many aviation authorities, e.g., the US-based \ac{FAA} and the EU-based \ac{EASA}, mandated remote identification of drones through dedicated \ac{RID} rules~\cite{remote_id},~\cite{belwafi2022_access}. According to such regulations, drones should constantly broadcast on the wireless channel RID messages containing a unique identifier, the drone's current location (latitude, longitude, altitude) and velocity, the location of the corresponding \ac{CS}, a time mark, and an emergency status indication. RID messages should be emitted no less than once a second in plain text, using either WiFi or BLE---when such communication technologies are unavailable onboard, the drone should be equipped with dedicated external RID modules.

Despite solving drones' accountability issues, RID regulations introduce confidentiality and privacy issues. Adversaries can track drones using a (network of) passive receiver(s). In turn, such tracking can lead to the leakage of private information about the drone, its pilot, and the drone-based business. For instance, the source location can be traced back to the location of warehouses, the flight destination can lead to customers' private location disclosure, and the track can reveal IP-protected information about the algorithm used for navigation~\cite{wisse2023_iotj}.  \textcolor{black}{A real-life example in this direction is the leakage occurred in October 2022, where a massive data leak exposed over 80,000 drone IDs and 90 million drone-monitoring logs from DJI AeroScope devices~\cite{aeroscope}.}
A few works investigated privacy issues connected with RID, mainly considering the identity of the drone and collision avoidance strategies~\cite{wisse2023_iotj},~\cite{tedeschi2022_tdsc}. However, to preserve full compliance with RID regulations, at the time of this writing, no works considered drone's location privacy. The mentioned problem shares similarities with privacy issues affecting users interacting with \acp{LBS}, where users would like to get suggestions based on their current location while not allowing \acp{LBS} providers to track them. However, providing location privacy to RID-enabled drones further complicates the problem, due to the required frequent information broadcast that makes the disclosed locations highly correlated in time and space. 
Moreover, compared to users' devices (smartphones, laptops), drones may not be persistently connected to the Internet. Also, drones can be more constrained in terms of processing, memory, communication, and energy availability, calling for very lightweight approaches. Overall, although we know from the literature that any location obfuscation mechanism is bound to cause privacy loss when used repeatedly~\cite{chatzikokolakis2014_pets}, a solution specifically tailored to the use cases of remotely-piloted RID-enabled drones is currently unavailable.

{\bf Contribution.} In our work, we design Obfuscated Location disclOsure for RID-enabled drones (\sol), a framework that allows RID-compliant remotely-piloted drones to protect as much as possible their current location while allowing relevant location-based services (using such location) to keep working reliably and efficiently. Our solution integrates and extends the mechanism proposed by Xiao et al. in~\cite{xiao2015_ccs}, making it suitable for running at runtime on drones while fulfilling all the requirements imposed by RID. We also extend the RID regulation through encrypted location reports that relevant authorities can use to disclose the drone location, if necessary. We implement our proposed solution on two \acp{PoC} devices, using a general-purpose laptop and a constrained embedded system \ac{RPI}, featuring similar hardware constraints to medium-end drones, and release the code of our implementation as open-source~\cite{code}. Through an extensive experimental measurement campaign, we show that \sol\ can run on relevant platforms by requiring less than $0.16$~s and $0.0236\%$ of energy--- considering the energy available on a DJI Mini 2. We also introduce three among the most relevant real-world use cases involving the usage of the drones' location and extract utility metrics to determine the impact of the disclosure of an obfuscated location on the quality of the services offered. Overall, we evaluate the trade-off between drones' location privacy and utility, showing the possibility of letting the two requirements coexist meaningfully and effectively.

Note that this work extends and completes our preliminary study in~\cite{brighente2022_ares} through the following contributions.
\begin{itemize}
    \item We replace the basic Geo-Indistinguishability method used in~\cite{brighente2022_ares} with a solution inspired by the proposal by Xiao et al. in~\cite{xiao2015_ccs}, specifically conceived for time-correlated location disclosures, modified on purpose to fit the scenario of RID-enabled drones.
    \item We implement the proposed solution on two \ac{PoC} platforms and release the code as open-source~\cite{code}.
    \item We evaluate the overhead of our solution on real drones' trajectory data in terms of execution time, memory, bandwidth, and energy consumption.
    \item We apply \sol\ to three relevant real-world use cases in the content of drone-based location-based services. We then show the trade-off existing between location privacy and utility in our proposed solution.
\end{itemize}

{\bf Roadmap.} This paper is organized as follows. Sec.~\ref{sec:preliminaries} introduces the preliminaries; Sec.~\ref{sec:scenario_description} describes the scenario, adversarial model, use cases, and requirements; Sec.~\ref{sec:related_work} discusses related work and evaluates their match to our requirements; Sec.~\ref{sec:core_proposal} introduces \sol; Sec.~\ref{sec:security_analysis} provides security considerations; Sec.~\ref{sec:impl} introduces our \ac{PoC}; Sec.~\ref{sec:evalution_and_results} discusses our extensive performance assessment; Sec.~\ref{sec:discussion} discusses the results; and finally, Sec.~\ref{sec:conclusion_future_work} concludes the paper.
\section{Preliminaries}
\label{sec:preliminaries}


\subsection{Remote ID}
\label{subsec:remote_id}
Remote ID (\ac{RID}) is a standard regulation first published in 2021 by the \ac{FAA}~\cite{remote_id}, whose goal is to enhance the accountability of \ac{UAV} operations and use the airspace efficiently while prioritizing public safety. This regulation forces almost all \acp{UAV} to emit wireless messages at least once each second, reporting their unique identity and location, the pilot's (\ac{CS}) location, the timestamp and emergency status. 
The regulation defines two \ac{RID} modes, i.e., Network \ac{RID} and Broadcast \ac{RID}. In Network \ac{RID}, the UAV delivers unicast \ac{RID} messages to a third-party service provider over the Internet, possibly relaying such messages through the \ac{CS}. In Broadcast \ac{RID}, the UAV broadcasts \ac{RID} messages using Wi-Fi or Bluetooth so that any compatible receiver nearby can receive such information. \acp{UAV} not equipped with a compatible transceiver must integrate external RID modules provided by relevant companies. This manuscript focuses explicitly on the Broadcast \ac{RID} mode, as this scenario presents unique challenges in realistic operational conditions.
The \ac{RID} rule applies to \acp{UAV} weighing more than $250$~g anywhere except FAA-Recognized Identification Areas, explicitly listed in the regulation. \ac{RID} does not specify any security measures applying to the delivered messages, leaving them free to the implementation. Thus, by default, RID messages are delivered in plaintext. 
Also, \ac{RID} does not provide specific techniques for deploying \ac{RID} in real-world scenarios. The \ac{DRIP} \ac{WG} of the IETF takes such gaps into account and is currently working on deployment and security requirements connected to \ac{RID}~\cite{drip}. Throughout several documents and \acp{RFC}, the \ac{WG} defines the network architecture for RID-compliance deployments, requirements and necessary protocols~\cite{rfc9153},~\cite{rfc9434}. In this work, we align the network architecture, entities, and design requirements to the ones defined by \ac{DRIP}, so creating a standard-compliant solution and smooth the possible integration of our solution into standardization activities. 

\subsection{Public Key Encryption}
\label{subsec:ecies}

\ac{PKE} techniques allow encrypting a plaintext message $m$ using the public key of the recipient $pk_R$ to a ciphertext $c$, such that the recipient, in possession of the corresponding private key $sk_R$, is the only party able to decrypt the ciphertext and obtain the plaintext.
\begin{definition}
A public key encryption algorithm $PKE$ includes the following algorithms:\\
$PKE.KGen(1^k)$: on input a security parameter $k$, it outputs a secret decryption key $sk$ and a public encryption key $pk$.\\
$PKE.Enc(pk_R,m)$: on input a plaintext message $m$ and a public key $pk_R$, it outputs a ciphertext $c$. \\
$PKE.Dec(sk_R,c)$: on input a ciphertext $c$ and a private decryption key $sk_R$, it outputs the corresponding plaintext $m$.
\end{definition}

In this manuscript, we use the well-known \ac{ECIES} as a \ac{PKE} technique~\cite{ecies_standard}. \textcolor{black}{Compared to symmetric encryption techniques, public-key encryption schemes like \ac{ECIES} allow reducing storage requirements and minimizing the secrets involved, thus easing system management.}\\\\
\section{Scenario, Use Cases, and Requirements}
\label{sec:scenario_description}


\subsection{Scenario and assumptions} 
\label{subsec:scenario_and_assumptions}

\textcolor{black}{We consider a \ac{RPAS}, compliant with the RID regulation}. Thus, it integrates a \emph{RID Module} that wirelessly broadcasts \ac{RID} messages from takeoff to shutdown. As discussed in Sec.~\ref{subsec:remote_id}, \ac{RID} messages contain a unique identifier of the UAV $UID$, the UAV's disclosed location data (latitude, longitude, and altitude) $\mathbf{z}$, the UAV's current velocity $\mathbf{v}$, the \ac{CS}'s current location (latitude, longitude, and altitude) $\mathbf{x_{CS}}$, a timestamp $t$, and the emergency status indication $em_t$. As per the regulation, we consider that the UAV has to broadcast a \ac{RID} message at least once a second, reporting every time an updated location. We consider \acp{UAV} broadcasting \ac{RID} messages via WiFi, so as to maximize transmission range while keeping compliance with the RID rule. \textcolor{black}{Consider that WiFi transceivers installed on \acp{UAV} can reach a range of a few kilometers, guaranteeing sufficient coverage~\cite{sciancalepore2022_acsac}}. We consider UAVs carrying a {\em \ac{GNSS} Module} to obtain the current location $\mathbf{x}$, e.g., a \ac{GPS} receiver. Moreover, we consider UAVs with enough computational power and energy to run basic cryptography algorithms, such as a \ac{PKE} scheme, through an {\em Encryption Module}. Based on the considered use case (see Sec.~\ref{subsec:use_cases}), the \ac{UAV} might operate in scenarios with an intermittent or unavailable Internet connection. Therefore, our solution (see Sec.~\ref{sec:core_proposal}) does not require the existence of a persistent connection of the \ac{UAV} to any \ac{CSP}. Although some use cases discussed below might describe the connection of \acp{UAV} to external services, such connection is never required to execute our solution. We also envision the presence of a dedicated process, namely the {\em Obfuscation Module}, aimed at generating an obfuscated location according to our solution. Onboard the UAV, for some use cases, we also consider the presence of a {\em Communication Module}, aimed at establishing mutual communication with other entities. This module can be installed by the owner of the UAV, based on the envisioned use case. 
As for the \ac{CS} location, we consider it is either already stored on the UAV or updated thanks to real-time communication with the \ac{CS}.\\
We also consider one or more wireless receivers, which we refer to as \emph{observers}, in line with the standard \ac{DRIP} architecture. Observers are entities that can receive \ac{RID} messages on a compatible RF interface. As defined in RFC 9153~\cite{rfc9153}, we consider multiple types of observers, namely General Public Observers, and Public Safety Observers. Public Safety Observers are part of a more extensive sensing infrastructure focusing on public safety and security. General Public Observers can be regular users receiving \ac{RID} messages without relationships with specific sensing infrastructures.
\\
Our system model also includes a \ac{TTP}, responsible for storing identifiers and cryptographic materials used to secure \ac{RID} messages. As per the architecture defined by \ac{DRIP}, we refer to the TTP as the \ac{USS}. Storage and retrieval of information from the \ac{USS} can be done through dedicated registries. In line with \ac{DRIP}, we distinguish between public and private registries~\cite{drip}. Using the \acp{UAV}' unique identifier, anyone can use the public registry to retrieve public information. Conversely, only authorized parties can use the private registry to retrieve confidential data. The TTP is always available online, so parties can interact with it if required.  

In our work, we want to design a location obfuscation mechanism to protect \acp{UAV}' actual location when disclosing RID messages, despite the presence of time correlation in the disclosed locations.
In Sec.~\ref{subsec:use_cases}, we provide a few use cases relevant to our investigation. We report the main notation used in the manuscript in the Appendix (supplementary materials).


\subsection{Adversary Model} 
\label{subsec:adversary_model}

\textcolor{black}{The adversary considered in this work is a static passive eavesdropper, namely $\epsilon$, who aims to track the UAV based on the disclosed \ac{RID} messages.} As \acp{UAV} deliver RID messages in plaintext, $\epsilon$ can achieve this aim by simply using a receiver compatible with the communication technology (WiFi). Using the retrieved location information, the attacker can track the UAV and deduct private information about the business the UAV is used for and the pilot. Such information can be further exploited to take a UAV down, capture it and steal any carried goods or private data stored onboard. All such actions could damage the UAV, the business using the \acp{UAV}, and potentially the safety of the people involved. 
\textcolor{black}{In line with relevant literature~\cite{wisse2023_iotj}~\cite{tedeschi2023_acsac}, our work considers honest UAVs. We intentionally do not consider scenarios where the pilot purposely turns off \ac{RID} or tamper with the RID module to carry out unauthorized actions, as this would involve malicious UAVs, which are not in the scope of our paper nor in that of the RID regulation. On the contrary, our work intends to protect the privacy of RID-compliant \acp{UAV} and the business using the \ac{UAV} from tracking and other threats, to finally unleash the potential behind their usage in various application domains.}\\

\subsection{Use Cases} 
\label{subsec:use_cases}

In the following, we describe three real-world use cases motivating the deployment of our solution.

{\bf Use Case 1: Unintentional Invasion of No-Fly Zone.} 
We consider a legitimate UAV operating near an \ac{NFZ}.
An \ac{NFZ} is an area where UAVs are not allowed to fly, e.g., around \acp{CI} such as airports and power distribution grids. We consider that the \ac{CI} operator maintaining the site deploys one or more observers to monitor the area of the \ac{NFZ} for UAV invasions (among other factors)---based on \ac{DRIP}, such observers are \emph{public safety observers}. Based on the deployed observers' area coverage, the system's overall reception range could extend outside the boundaries of the \ac{NFZ}. We define the portion of the coverage area outside of the \ac{NFZ} as the \ac{WA}. In principle, UAVs are allowed to fly in the \ac{WA}. When receiving RID messages, the observers use the disclosed location to check if the UAV is invading the \ac{NFZ}. If this is the case, they take action and use the disclosed location to take the \ac{UAV} down (e.g., by jamming or physically shutting it down--through weapons or nets)~\cite{sciancalepore2023_iotj}. 
Consider \acp{UAV} disclosing an obfuscated location through RID messages rather than the actual one, and approaching the \ac{NFZ}. Several situations may occur: (i) the disclosed location lies outside of the \ac{NFZ}, and also the actual location lies outside of the \ac{NFZ}--this is a \ac{TN}; (ii) the disclosed location is inside of the \ac{NFZ}, but the actual location lies outside of the \ac{NFZ}---this is a \ac{FP}; (iii) the disclosed location lies outside of the \ac{NFZ}, but the actual location is inside the \ac{NFZ}---this is a \ac{FN}; and finally, (iv) the disclosed location is inside the \ac{NFZ}, and also real location lies is inside the \ac{NFZ}---this is a \ac{TP}. Moreover, when detecting a disclosed location inside the \ac{NFZ}, the \ac{CI} operator has to know the actual location to verify the invasion and possibly take action. However, the actual location of the UAV has not been disclosed. 
In turn, such situations would have a significant impact, as \acp{UAV} inadvertently invading the NFZ might cause severe security, safety and operational issues~\cite{wigchert2025_comnet}. 
Therefore, protecting \acp{UAV}' location privacy while allowing \ac{CI} providers to detect invasions and take \acp{UAV} down is challenging and requires the design of a tailored solution. We remark that, for this use case, we consider benign \acp{UAV} unintentionally invading the \ac{NFZ}. Malicious \acp{UAV}, intentionally invading the NFZ to cause damage, cannot be detected by relying only on RID messages, since malicious \ac{UAV} operators might easily stop transmitting RID messages. Conversely, the aim of this use case is to investigate the trade-off between location privacy and utility for benign \acp{UAV}.


{\bf Use Case 2: Discovery of the closest Charging Station.} Consider a \ac{SP} managing a charging support system for \acp{EV} consisting of a group of \emph{Charging Stations}. The SP distributes several charging stations in a given area and allows any \acp{EV}, including \acp{UAV}, to recharge by paying a fee. We consider that charging stations can receive \ac{RID} messages and communicate with a central server online. Considering all the charging stations that received RID messages from a given UAV, being the only entity knowing the exact location of the control stations, the central server can determine the nearest control station to the UAV disclosed location. A control station can communicate with \acp{UAV}, e.g., through a secure communication link.

Consider a UAV on a mission, implementing location obfuscation to achieve location privacy. During the mission, the UAV needs to find the closest charging station to recharge its battery. 
As the UAV discloses an obfuscated location through RID messages, the central server of the charging support system can compute the nearest charging station only based on the disclosed location, not the actual one (this is never disclosed). As a result, the system might not be able to suggest to the UAV the charging station closest to the UAV's actual location. 
A sub-optimal choice can cause extra distance to be travelled by the \ac{UAV} and more energy consumption, possibly leading to battery drain, energy exhaustion and mission failure.
Therefore, protecting \acp{UAV}' location privacy while allowing \acp{SP} to provide service is challenging and requires the design of a tailored solution. We remark that the aim of this use case is to investigate the trade-off between location privacy and utility for \ac{UAV}-based \ac{LBS}. Real charging stations might also use a different protocol, e.g., broadcasting their own location so to allow UAV to determine the best charging station autonomously. However, since no trade-off would be involved in such a scenario, we do not consider it for our analysis.


{\bf Use Case 3: \ac{UAV}-as-a-Service.} 
We consider \acp{UAV} deployed by a telecommunication \ac{SP} and providing a service to users nearby, in line with the emerging \ac{DaaS} scenario~\cite{george2023_sacmat}. In particular, we assume such \acp{UAV} are equipped with a \ac{SIM} card that allows them to connect to the mobile cellular network, relaying requests coming from active users. Thus, users without Internet connectivity in remote regions can connect to such \acp{UAV}, establish a secure connection, and make phone calls or access services available online. 

Consider that the \acp{UAV} implement location obfuscation to achieve location privacy. Thus, when users would like to use a service provided by the \ac{SP} via the \acp{UAV}, they connect to the \ac{RF} interface (WiFi), detect all UAVs in a given area and pick the one whose disclosed distance (via RID) is the least from the user location, assuming the closest \ac{UAV} is the one providing the best signal quality. 
As the UAVs disclose obfuscated locations, the user might make a sub-optimal choice, i.e., pick the UAV whose actual location is not the least. Thus, the signals emitted by the user and the UAV have to travel extra distance than the minimum, with consequent degradation of the achievable Quality of Service (QoS). 
Therefore, protecting a \ac{UAV}'s location privacy while allowing a \ac{SP} using such \acp{UAV} to provide reliable and efficient location-based services is challenging and requires the design of a tailored solution.

Note that users of DaaS applications might also use other parameters to pick the best UAV (e.g., the Received Signal Strength-RSS). However, channel quality indicators such as the RSS notoriously exhibit fast-fading phenomena that make distance estimation highly error-prone~\cite{tedeschi2021_spaccs}.


\subsection{Requirements}
\label{subsec:requirements}

From our previous discussion, we can extract a list of requirements for our proposal. These requirements can serve as a baseline to evaluate the applicability of different state-of-the-art techniques to our problem (see Sec.~\ref{sec:related_work}) and to assess the overall system performance (see Sec.~\ref{sec:evalution_and_results}).
\begin{itemize}
\label{state-of-the-art-requirements}
    \item {\bf R1: Location Obfuscation.} The latitude, longitude, and altitude reported by the UAV through \ac{RID} messages must be obfuscated from takeoff to shutdown, i.e., its value should never be disclosed in plaintext. 
    \item {\bf R2: Trajectory Privacy.} A passive eavesdropper must not be able to use temporal correlation in the disclosed locations to extract the real UAV location from the obfuscated location released in RID messages.
    \item {\bf R3: No Persistent Internet Connection on \acp{UAV}.} The method to obfuscate the location must work on the UAV independently from the availability of a persistent Internet connection onboard.
    \item {\bf R4: Maximum Messages Generation Time.} To keep compliance with the RID regulation, the time necessary to generate a \ac{RID} message including an obfuscated location must not exceed $1$~second.
    \item {\bf R5: Maximum Communication Overhead.} The communication overhead resulting from the introduction of location obfuscation should not lead to message fragmentation, due to the low reliability of broadcast communications and limited energy budget. Thus, \ac{RID} messages must not exceed the \ac{MTU} of WiFi, i.e., $2,304$~B~\cite{ieee2012}.
\end{itemize}
\section{Related Work}
\label{sec:related_work}

Several contributions in the current literature investigate location privacy issues and protection mechanisms in the context of vehicular networks. 
Some contributions provide location privacy via anonymity, as it can make tracking more difficult. One way to provide location privacy via anonymity is via pseudonyms~\cite{svaigen2022_iotmag},~\cite{tedeschi2021_acsac}, i.e., short-term identifiers that keep changing on a (regular) time basis. Note that if an adversary knows the location and the time before the pseudonym changes, they can link the old pseudonym to the new one, breaking location privacy. \\
In the avionic domain, Sampigethaya et al.~\cite{Sampigethaya2009_dasc} propose to create areas dedicated to pseudonym swap where aircraft do not transmit messages, making it harder for a passive adversary to track aircraft based on wireless transmissions. This strategy cannot apply to our scenario, as \ac{UAV}s should comply with RID. Svaigen et al.~\cite{svaigen2022_icc} propose creating dummy data and mobile entities. When applied to our context, these techniques would break compliance with RID and cause extra energy consumption.

Location privacy can also be achieved via location obfuscation, i.e., perturbating the location disclosed to other parties to prevent precise location tracking~\cite{jiang2021_csur}. Geo-Indistinguishability is a well-known location obfuscation method proposed by Andres et al.~~\cite{andres2013_ccs}, which protects the user's location over multiple location disclosures. Such an approach also preserves location utility to \acp{LBS}. Geo-Indistinguishability was designed for low-frequency location disclosure with no time correlation. The achievable location privacy decreases when sending messages frequently \cite{mendes2020_pets}, as passive attackers can reveal the actual location through inference attacks \cite{xiao2015_ccs}. Thus, many contributions extend Geo-Indistinguishability to face such inference attacks in various scenarios. 
Chatzikokolakis et al.~\cite{chatzikokolakis2014_pets} focus on supporting location traces for Geo-Indistinguishability. Their proposal reduces the computation cost by reusing noise, providing enough privacy for an entire day. However, they only used a limited number (30) of daily queries. 
Hua et al.~\cite{hua2017_tifs} focus on reporting obfuscated locations in rapid succession. The proposed mechanism can work in both a basic and advanced mode. 
Such proposal assumes an external trusted service computes the obfuscated location, so it does not apply to our scenario. As discussed by Svaigen et al. in~\cite{svaigen2022_iotmag}, another option includes not revealing the exact location but the one of an entity somewhere in an area. The security and utility of such a system depend on the size of the site. Another option discussed by Svaigen et al. in~\cite{svaigen2022_iotmag} is to use a protocol that works on a node-to-node basis, which is also not an option if the \ac{UAV} is not working in a swarm. Sampigethaya et al.~\cite{Sampigethaya2009_dasc} also propose using cooperative schemes, which is inapplicable to our problem.
Other proposed mechanisms either require time and energy-consuming computations before using the system, such as the one by Zhang et al.~\cite{zhang2019_tifs}, or have a chance to release the actual location, such as the one by Xiong et al. \cite{xiong2019_ijdsn}, which cannot fulfil our requirements. 
Xiao et al.~\cite{xiao2015_ccs, xiao2017_vldb} specifically focus on the temporal correlation between location reports. They use a $\delta$-location set to protect the areas where the entity is most likely and develop an obfuscation method, namely \ac{PIM}, which supports frequent location releases through a lightweight location obfuscation mechanism. Cao et al.~\cite{cao2019_icde} combine Geo-Indistinguishability with the proposal by Xiao et al.~\cite{xiao2015_ccs} and focus on obfuscating specific events, i.e. ``moving between two specific places". Through this strategy, an adversary cannot tell if an event has occurred. However, such a method cannot protect trajectory privacy.
In the \ac{UAV} context, a recent work by Enayati et al. considered the application of Laplacian Geo-Indistinguishability to protect UAV location privacy in goods delivery applications and \ac{UAV}-IoT applications~\cite{enayati2023_iotj}. However, the authors do not consider compliance with \ac{RID} regulations.
To sum up, we cross-compare in Tab. \ref{tab:technique_comparison} the described techniques and evaluate if and how they fulfil our requirements (Sec.~\ref{subsec:requirements}).


\begin{table}[htbp]
\centering
\caption{Comparison of state-of-the-art techniques for location privacy regarding our requirements of interest discussed in Sec.~\ref{subsec:requirements}. The symbol \newmoon\ indicates that the paper fulfils the requirement, and \fullmoon\ indicates that the paper does not address the requirement. In contrast, \LEFTcircle\ indicates partial fulfilment of the requirement (explanation in text).}
\label{tab:technique_comparison}
\begin{tabular}{l|lllll}
{\bf Ref.} & {\bf R1} & {\bf R2} & {\bf R3} & {\bf R4} & {\bf R5}\\ \hline
\cite{chatzikokolakis2014_pets} &\LEFTcircle&\fullmoon&\newmoon&\fullmoon&\newmoon\\\hline
\cite{xiao2015_ccs} &\LEFTcircle&\newmoon&\newmoon&\fullmoon&\newmoon\\\hline
\cite{brighente2022_ares}
&\newmoon&\LEFTcircle&\newmoon&\fullmoon&\newmoon\\ \hline
\cite{svaigen2022_iotmag} &\LEFTcircle&\fullmoon&\newmoon&\fullmoon&\newmoon\\\hline
\cite{Sampigethaya2009_dasc} &\LEFTcircle&\newmoon&\fullmoon&\fullmoon&\newmoon\\\hline
\cite{andres2013_ccs}&\LEFTcircle&\fullmoon&\newmoon&\fullmoon& \newmoon\\\hline
\cite{hua2017_tifs} &\LEFTcircle&\newmoon&\fullmoon&\newmoon&\newmoon\\\hline
\cite{zhang2019_tifs} &\LEFTcircle&\newmoon&\newmoon&\fullmoon&\newmoon\\\hline
\cite{xiong2019_ijdsn} &\LEFTcircle&\fullmoon&\newmoon&\fullmoon&\newmoon\\\hline
\cite{cao2019_icde} &\LEFTcircle&\fullmoon&\newmoon&\fullmoon&\newmoon\\ \hline
\cite{enayati2023_iotj}
&\LEFTcircle&\LEFTcircle&\newmoon&\fullmoon&\newmoon\\ \hline
\textbf{OLO-RID}&\newmoon&\newmoon&\newmoon&\newmoon&\newmoon \\ 
\end{tabular}
\end{table}
Note that none of the available solutions fulfils all our requirements, at the same time, motivating our work. 
The proposals which fulfil most of our requirements are the ones by Xiao et al.~\cite{xiao2015_ccs},~\cite{brighente2022_ares},~\cite{hua2017_tifs}, and~\cite{zhang2019_tifs}. Some approaches cannot be improved to deal with trajectory privacy (e.g.,~\cite{brighente2022_ares}), while others cannot run on \acp{UAV} due to heavy computational demands (\cite{hua2017_tifs} and~\cite{zhang2019_tifs}). As for the proposal by~\cite{xiao2015_ccs} et al., it has not been conceived for 3D scenarios, it has not been tailored to constrained \acp{UAV}, and no considerations about their utility in \ac{UAV}-based scenarios have been provided. We integrate such an approach into our proposed solution as it achieves the lower bound of location privacy in equivalent 2-D scenarios. We will describe the integration process, extension, and evaluation in the following sections.
\section{\sol}
\label{sec:core_proposal}

This section provides the details of \sol. We first describe the location obfuscation procedure in Sec.~\ref{sec:pim_changes}. Then, we describe the operations occurring on the \ac{UAV} in Sec.~\ref{subsec:protocol_flow} and, finally, provide the details of the additional procedures required for each of the use cases in Sec.~\ref{sec:proto_usecases}.

\subsection{Location Obfuscation on RID-enabled \acp{UAV}}
\label{sec:pim_changes}

Our solution, namely \sol, leverages the mathematical model of the PIM approach proposed by Xiao et al. in~\cite{xiao2015_ccs}. We first review the original approach and then discuss its limitations and our contributions.

\textbf{Introduction to PIM.} For ease of discussion, consider a two-coordinate system (latitude-longitude). 
We divide the area around the moving entity into multiple cells $s_1, s_2, ..., s_{N}\in S$. Based on previous travel data of the mobile entity or public data about such a journey, PIM requires us to create a transition matrix $\mathbf{M}$, holding the probability that the mobile entity moves from one cell to another. We denote $M_{ij}$ as the probability that the mobile entity moves from cell $i$ to cell $j$. We also consider a probability array $\mathbf{p}$, indicating the probability that the moving entity is in the cell $s_i$. As an example, if the mobile entity has a uniform probability of residing in cells 1, 2, 3, and 4, then $p=[0.25,0.25,0.25,0.25,0,0,...]$. We define $p_t^-$ as the prior probability at timestamp $t$, i.e., the probability that at time $t$, before a location release, the moving entity resides in a given cell. We also define $p_t^+$ as the posterior probability, i.e., the probability that at time $t$, after a location release, the moving entity resides in a given cell. According to PIM, we compute the prior probability as $p_t^- = p_{t-1}^+\mathbf{M}$. We can also compute the posterior probability according to Eq.~\ref{eq:posterior}, where $\mathbf{z}_t$ is the location released at time $t$, and $\mathbf{u}_t$ is the actual location.
\begin{equation} 
    \label{eq:posterior}
    p_t^+[i] = \frac{Pr(z_t|u_t = s_i)p_t^-[i]}{\sum_j Pr(z_t|u_t = s_j)p_t^-[j]}.
\end{equation}

For a timestamp $t$, given the prior probabilities of the mobile entity's location $p_t^-$ and the cells $s_i \in S$, we define the $\delta$-location set $\Delta X$ as per Eq.~\ref{eq:delta_location_set}.
\begin{equation} 
    \label{eq:delta_location_set}
    \Delta X_t = min\{s_i|\sum_{s_i} p_t^-[i] \geq 1 - \delta \} .
\end{equation}

Consider the actual location of the mobile entity $\mathbf{x}$. The PIM mechanism aims to generate the obfuscated location offset $\mathbf{z^{\#}}$, s.t. $\mathbf{z} = \mathbf{x} + \mathbf{z^{\#}}$. To this aim, PIM applies the following procedure. 
We compute the convex hull from the $\delta$-location set $\Delta X$. Given a set of points and drawing lines between each point, the convex hull is the smallest area not crossing any line \cite{xiao2015_ccs}. Denote a vertex of the convex hull as $v_i$. We denote $\Delta V$ as the set of all values $v_i - v_j$ for all possible $i$ and $j$. From this, we calculate the sensitivity hull $K$ of a convex hull of $\Delta V$. A sensitivity hull is a convex hull capturing the sensitivity between two releases from two instances, i.e., the difference between two queries if one piece of data has changed. From the sensitivity hull $K$, we sample points uniformly. These points are used in an isotropic transformation, resulting in the matrix $T$. To calculate the isotropic position from $K$, we get $\mathbf{K_I} = TK$. We sample from $\mathbf{K_I}$ a point uniformly at random, namely $\mathbf{z^{'}}$. To increase randomness, we use a Gamma distribution $\Gamma(3,\epsilon^{-1})$ to get $r$. 
We denote the new random location as $\mathbf{z^{*}} = r\cdot \mathbf{z^{'}}$. \textcolor{black}{Note that there is a small probability that the location $\mathbf{z^{*}}$ lies outside of the $\delta$-location set. In such cases, PIM uses a surrogate location $\Tilde{x}_t$, selected as the cell in $\Delta X$ closer to the actual location $\mathbf{x}$.} This location has to be transformed back to the original space $\mathbf{z^{\#}} = T^{-1}\mathbf{z^{*}}$. Finally, we can calculate our obfuscated location $\mathbf{z}$ using the real location $\mathbf{x}$ and $\mathbf{z^{\#}}$, thus we get $\mathbf{z} = \mathbf{x} + \mathbf{z^{\#}}$. Expanding the formula, we have that $\mathbf{z} = \mathbf{x} + rT^{-1}\mathbf{z^{'}}$. \\
\textbf{Extension of PIM.} As discussed in Sec.~\ref{sec:related_work}, the solution introduced by Xiao et al. in~\cite{xiao2015_ccs}, namely PIM, does not fulfill all our requirements natively, preventing its straightforward integration for our problem.
First, \ac{PIM} protects the location in a two-dimensional coordinate system. However, UAVs move across a three-dimensional space. RID messages still leak information when obfuscating only two dimensions, e.g., if the UAV is taking off or landing, leading to privacy issues. 
Thus, we extend the method by Xiao et al. to support location obfuscation in a 3-D space. 
\textcolor{black}{
Another downside of PIM when adopted in our context is the transition matrix, which contains a specific bounded area where the UAV is supposed to fly. When the UAV flies far outside these bounds, the computed obfuscated location is still based on a location within bounds, resulting in the computation of an obfuscated location very far from the actual location. In turn, such difference makes disclosed locations largely unrealistic. A naive solution could be uploading an extensive transition matrix on the UAV, so it likely always stays within bounds. However, a UAV is a constrained device with limited memory and computation power. Thus, loading an extensive transition matrix in memory is not an option. The computation time would also increase because of more extensive matrix computations.
To solve this issue, our solution uses a small transition matrix supporting the areas around the UAV in all three dimensions. The \ac{UAV} gets information about the prior probabilities from publicly accessible or personal flight data. This is indeed feasible, as \acp{UAV} likely operate in well-defined areas, with limited size. In some cases, e.g., autonomous \acp{UAV}, the precise travel path is even known in advance. However, in the more general case of \acp{RPAS}, a \ac{UAV} could move in any direction with the same probability. To generate the probabilities of a UAV flying from one cell to another, we initialise all the cells in the transition matrix to have the same probability $\frac{1}{b^3}$, in line with the logic of remotely piloted systems. Specifically, we check whether a cell near the UAV at timestamp $t$ is also present at timestamp $t-1$. If true, the cell is regularly updated as $p_t^{-} = p_{t-1}^+ \cdot M$. Otherwise, the value of $p_t^{-} = \frac{1}{b^3} \cdot M.$ }
Algo.~\ref{algo:entire_system} provides the pseudo-code of our modified location obfuscation technique. 
\begin{algorithm}
    \SetKwInOut{Input}{Require}
    \SetKwFunction{Algo}{Algorithm \ref{algo:pim}}
    \SetKwFunction{Pim}{Algorithm \ref{algo:entire_system}}
    \SetKwFunction{Perm}{Permute}
    \Input{$\epsilon_t, \delta, \mathbf{M}, \mathbf{p_{t-1}^+}, \mathbf{x_t^*}, o_{x}, o_{y}, o_{z}, \mathbf{S^{\#}_{t-1}}, b$}
    $\mathbf{o}$ = \Perm{$[0,1,-1]$}$[o_{x}, o_{y}, o_{z}]$\;
    $\mathbf{S^{\#}_t} = \mathbf{x_t}^* + \mathbf{o}$\tcp*{location offsets}
    $\mathbf{p_{\#}} = [\frac{1}{b^3}, ..., \frac{1}{b^3}]$\;
    \For{any two location $i, j$ in $S^{\#}_t$ and $S^{\#}_{t-1}$, respectively}{
        \If{$i \approx j $}{
            $p_{\#}[i] = p_{t-1}^+[j]$
        }
    }
    $\mathbf{p_t^-} \leftarrow \mathbf{p_{\#}} \cdot \mathbf{M}$\;
    \If{location needs to be released}{
        Construct $\Delta X_t$\tcp*{$\delta$-location set}
        \If{$\mathbf{x_t^*} \notin \Delta X_t$}{
            $\mathbf{x_t^*} \leftarrow \Tilde{x}_t$\tcp*{surrogate}
        }
        \tcp{release $\mathbf{z}$}
        $\mathbf{z} \leftarrow 
        \mathbf{x^*} + rT^{-1}\mathbf{z^{'}}$
        \;
        Derive posterior probability $p_t^+$ by Equation \ref{eq:posterior}\;
    }
    \Return{\Pim{$\epsilon_{t+1}, \delta, M, p_t^+, x_{t+1}^*, S^{\#}_t$}}
    \caption{Extended PIM approach for location obfuscation in \sol.}
    \label{algo:entire_system}
\end{algorithm}
We remark that the probability that the \ac{UAV} re-initializes the values of the prior probabilities to the default value ($\frac{1}{b^3}$) can be made arbitrarily small based on the knowledge of the area where the UAV has to fly and the autonomy of the battery deployed onboard.
\textcolor{black}{Finally, we integrate the extended mechanism discussed above into the \ac{RID} specification, defining the complete message format and protocol flow in a way that is compliant and applicable to RID-enabled devices, as well as backward compatible with RID receivers.}

\subsection{\sol\ Protocol flow}
\label{subsec:protocol_flow}

\begin{figure}[htbp]
\begin{center}
\includegraphics[width=8cm]{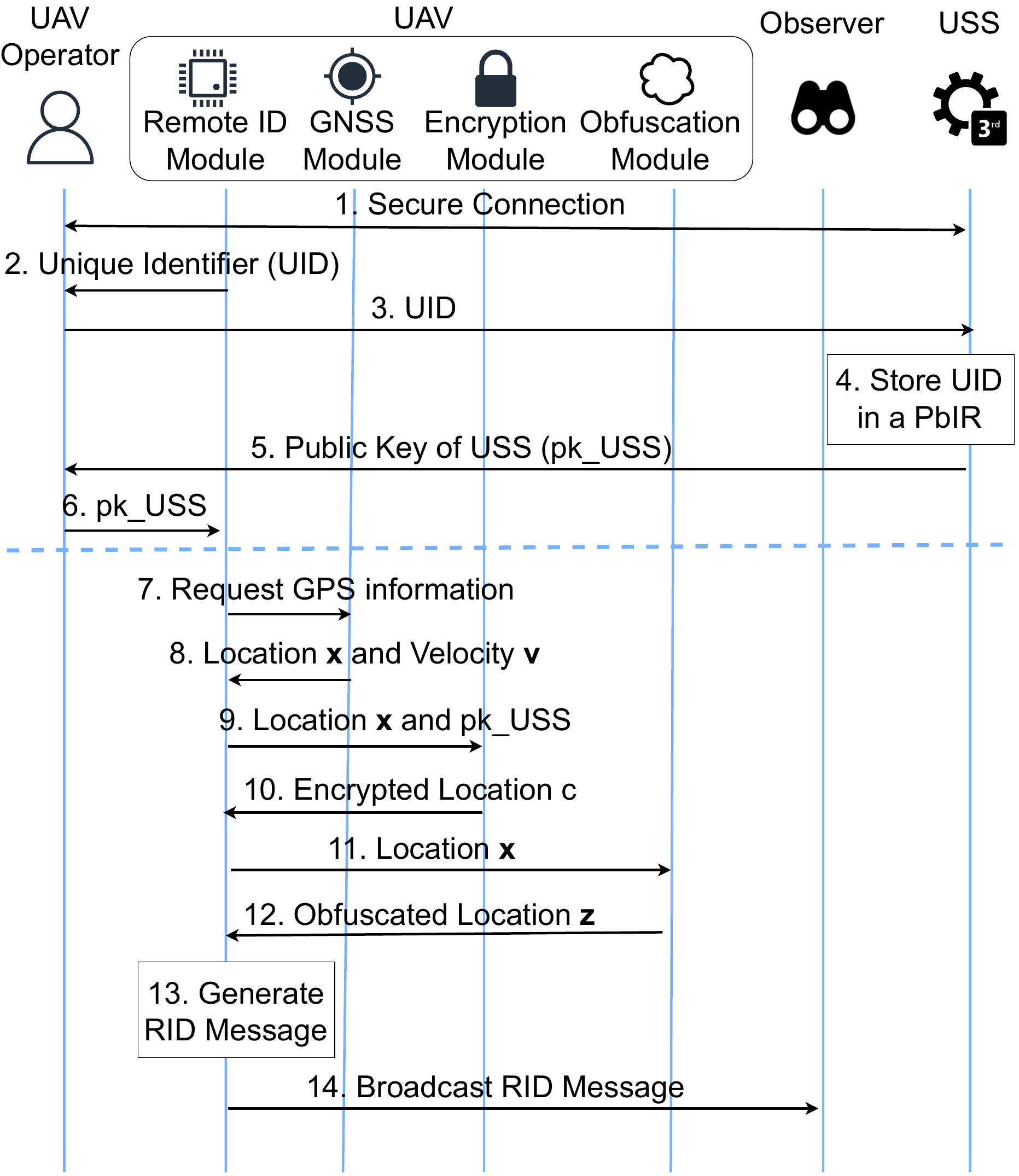}
\caption{Sequence diagram of the operations required for the \emph{Registration Phase} (upper part) and \emph{Runtime Phase} (lower part).}
\label{fig:seq_complete}
\end{center}
\end{figure}

\sol\ involves a \emph{Registration Phase}, executed before \ac{UAV} deployment, and a \emph{Runtime Phase}, executed during deployment. Fig.~\ref{fig:seq_complete} provides an overview of the operations required in such two phases, as described below.

\begin{itemize}
    \item[1] To start the \emph{Registration Phase}, the \textit{UAV operator} and the \textit{\ac{TTP}} establish a secure connection over the Internet, e.g., through TLS, to register the UAV. Such a registration only happens once for a given UAV.
    \item[2] The \textit{UAV operator} retrieves the \ac{UID} $UID$ from the \textit{UAV}, and specifically, from the \textit{RID Module}.
    \item[3] The \textit{UAV operator} delivers $UID$ to the \textit{\ac{TTP}} through the secure connection.
    \item[4] The \textit{\ac{TTP}} stores $UID$ in the \ac{PbIR}, for future use.
    \item[5] The \textit{\ac{TTP}} delivers to the UAV operator the public key $pk_{USS}$, through the secure connection.
    \item[6] The \textit{UAV operator} installs the key $pk_{USS}$ onto the \textit{UAV}, to be used later on in the {\em Runtime Phase}. The \emph{Registration Phase} ends here.
    \item[7] The \emph{Runtime Phase} is executed at runtime every time the UAV needs to create a new \ac{RID} message. In particular, the \textit{RID Module} on the UAV first requests the current location to the \textit{\ac{GNSS} Module} onboard.
    \item[8] The \textit{\ac{GNSS} Module} extracts the UAV's location $\mathbf{x}$ and velocity $\mathbf{v}$ and delivers them to the \textit{RID Module}.
    \item[9] The \textit{RID Module} on the UAV sends the location $\mathbf{x}$ and the public key $pk_{USS}$ to the \textit{Encryption Module}.
    \item[10] The \textit{Encryption Module} encrypts the location $\mathbf{x}$ using the ECIES scheme (Sec.~\ref{sec:preliminaries}) as $c = PKE.ENC(x_{lat}||x_{lon}||x_{alt}, pk_{USS})$, and sends it back to the \textit{RID Module}.
    \item[11] The \textit{RID Module} sends the location $\mathbf{x}$ to the \textit{Obfuscation Module}.
    \item[12] The \textit{Obfuscation Module} generates an obfuscated location $\mathbf{z}$ from the location $\mathbf{x}$, as $\mathbf{z} = PIM(\mathbf{x})$, and sends it back to the \textit{RID Module}.
    \item[13] The \textit{RID Module} generates a \ac{RID} message $m$ using the \ac{UID}, obfuscated location $\mathbf{z}$, velocity $\mathbf{v}$, \ac{CS}'s location $\mathbf{x_{CS}}$, the current timestamp $t$, the emergency status at time $t$ $em_t$, and the encrypted location $c$, as $m=UID||\mathbf{z}||\mathbf{v}||x_{CS}||t||em_t||c$.
    \item[14] The UAV broadcasts the \ac{RID} message over the wireless channel. 
\end{itemize}

\subsection{Application of \sol\ in Reference Use Cases}
\label{sec:proto_usecases}

In this section, we describe how the services connected to the use cases described in Sec.~\ref{subsec:use_cases} can use the location disclosed by the \acp{UAV} via \sol. 

\begin{figure}[htbp]
\begin{center}
\includegraphics[width=7cm]{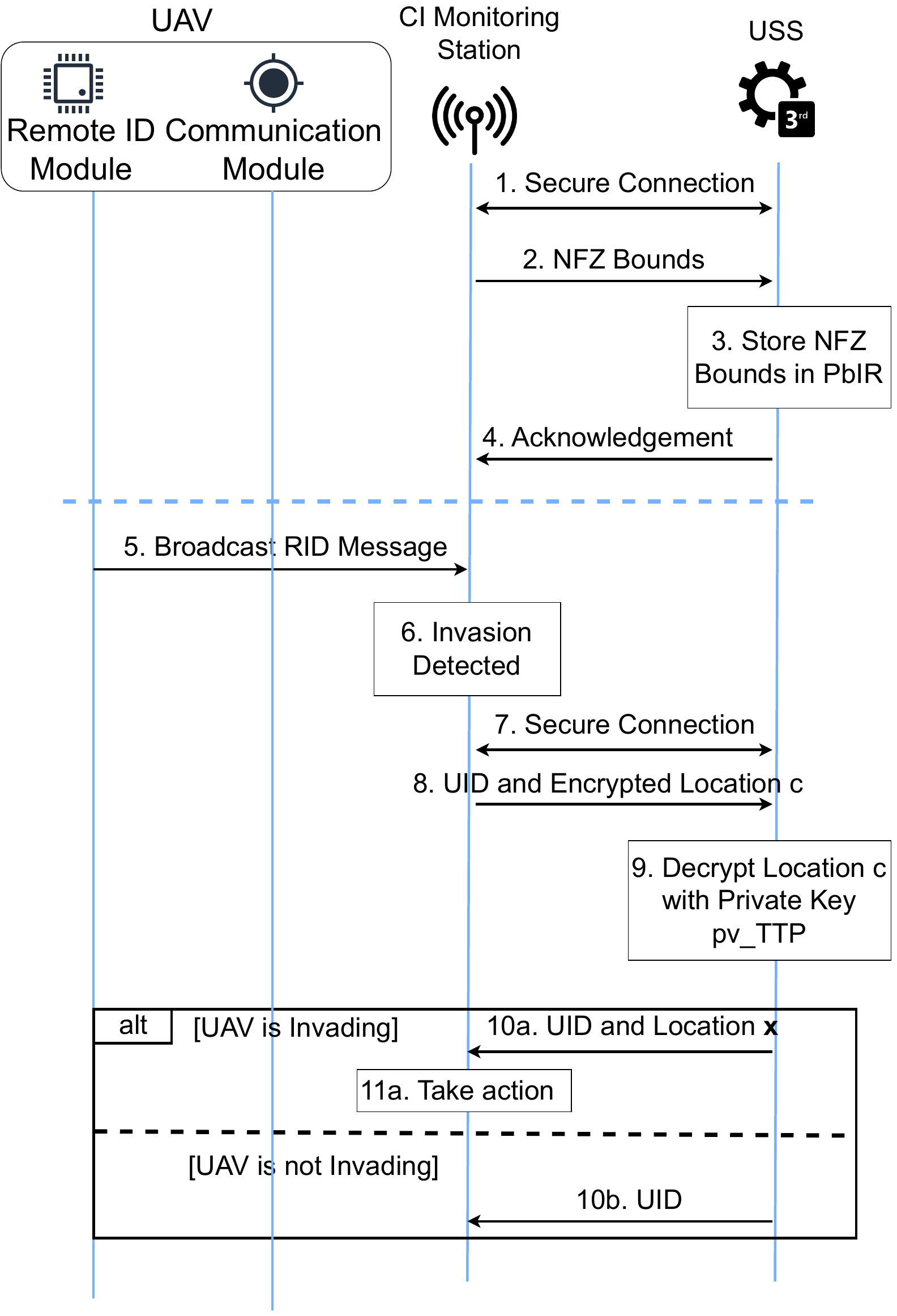}
\caption{Sequence diagram of the operations executed in Use Case \#1 to register the \ac{NFZ} (upper part) and handle a \ac{RID} message on the observer (lower part).}
\label{fig:seq_nfz}
\end{center}
\end{figure}

\textbf{Use Case 1: Invasion of No-Fly Zone.} Consider a \ac{RID}-equipped UAV is approaching a \ac{NFZ}, monitored by a {\em CI \ac{MS}}. We show the protocol flow in this use case in Fig.~\ref{fig:seq_nfz} while describing the steps below.
\begin{itemize}
    \item[1]  In the \emph{Registration Phase}, the \textit{\ac{CI} \ac{MS}} and the \textit{\ac{TTP}} establish a secure connection over the Internet, e.g., through TLS, to register the bounds of the \ac{NFZ} for further usage. Such a registration must only happen once for a given \ac{CI}.
    \item[2] The \textit{\ac{CI} \ac{MS}} delivers the \ac{NFZ} bounds to the \textit{\ac{TTP}} to register the \ac{NFZ}.
    \item[3] The \textit{\ac{TTP}} stores the received \ac{NFZ} bounds in a the \ac{PvIR} for further usage.
    \item[4] The \textit{TTP} delivers an acknowledgement message back to the \textit{\ac{CI} \ac{MS}}, to confirm bounds registration. This operation ends the \textit{\ac{NFZ} Registration Phase}.
    \item[5] At runtime, the \textit{UAV} broadcasts a \ac{RID} message through the {\em RID module}, as discussed above.
    \item[6] The \textit{\ac{CI} \ac{MS}} receives the \ac{RID} message and checks whether the obfuscated location is within the bounds of the \ac{NFZ}. The protocol flow described here continues only if the obfuscated location reported by the UAV is within the \ac{NFZ}.
    \item[7] The \textit{\ac{CI} \ac{MS}} and the \textit{\ac{TTP}} establish a secure connection over the Internet (e.g., via TLS) to exchange data about the UAV's location.
    \item[8] The \textit{\ac{CI} \ac{MS}} forwards the message received from the \textit{UAV} to the \textit{\ac{TTP}}.
    \item[9] The \textit{\ac{TTP}} decrypts the encrypted location report $c$ in the received message $m$, using its private key $sk_{USS}$ and the ECIES algorithm, as $\mathbf{x} = PKE_.DEC(c,sk_{TTP})$, so retrieving the \textit{UAV}'s actual location. The following actions depend on whether the \textit{UAV} is actually invading. When the UAV invades the \ac{NFZ}, we report the steps below with an `a'. Otherwise, we report them with a `b'.
    \item[10a] If the UAV is actually invading the NFZ, the \textit{\ac{TTP}} delivers the UID and the plaintext location of the \textit{UAV} $\mathbf{x}$ to the \textit{\ac{CI} \ac{MS}}.
    \item [11a] The \textit{\ac{CI} \ac{MS}} takes action against the \textit{UAV}.
    \item[10b] If the UAV is not invading, the \textit{\ac{TTP}} sends only the UID to the \textit{\ac{CI} \ac{MS}}. It never reveals the plaintext location, protecting the UAV's location privacy.
\end{itemize}

\begin{figure}[htbp]
\begin{center}
\includegraphics[width=7cm]{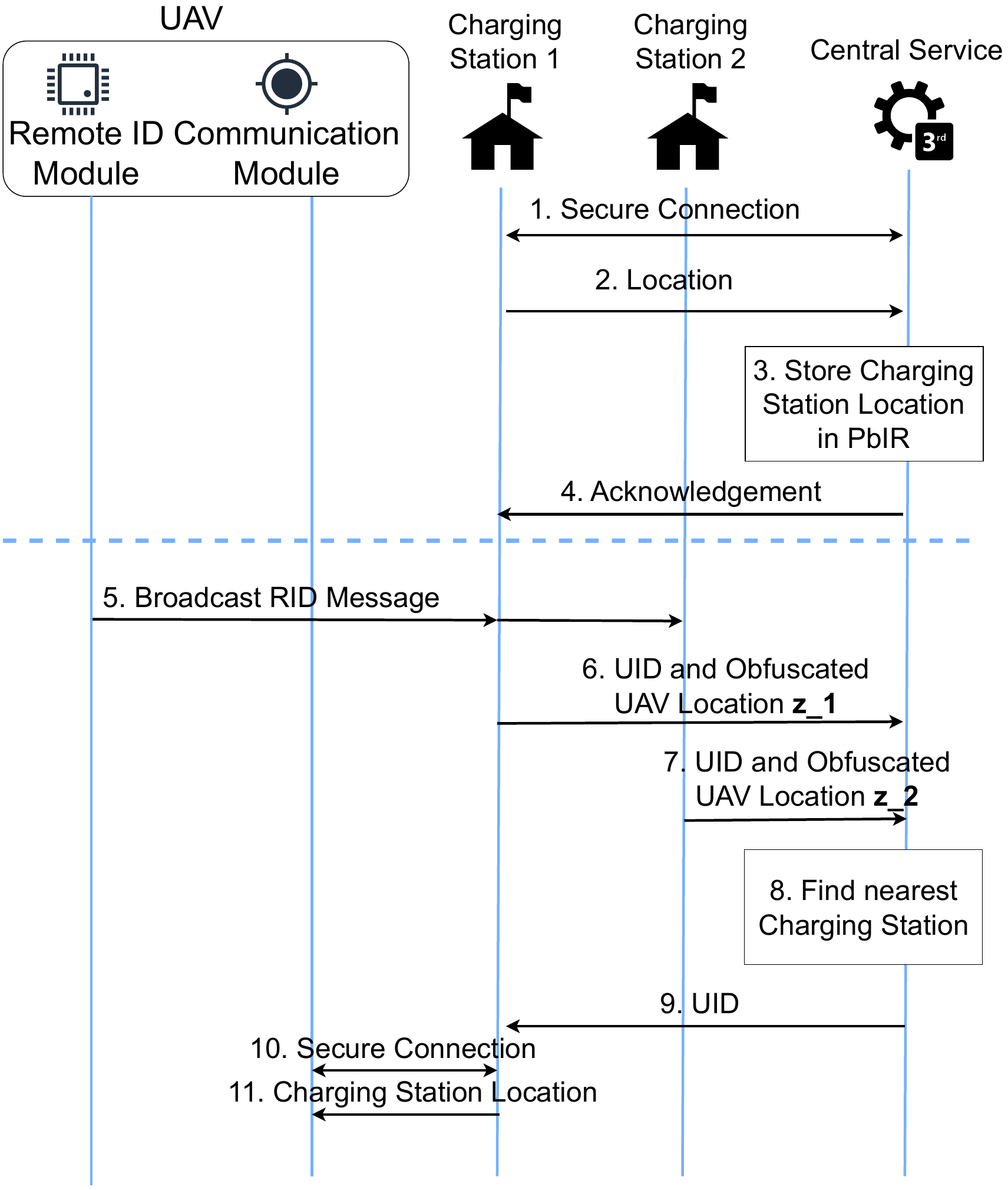}
\caption{Sequence diagram of the operations executed in Use Case \#2 to register charging stations (upper part) and handle a \ac{RID} messages on the observers (lower part).}
\label{fig:seq_charging_station}
\end{center}
\end{figure}

\textbf{Use Case 2: Discovery of the closest Charging Station.} Consider a RID-enabled UAV looking for the closest charging station. We show the protocol flow for this use case in Fig.~\ref{fig:seq_charging_station} while we describe the required steps below.
\begin{itemize}
    \item[1] In the \emph{Registration Phase}, each \textit{Charging Station $j$} and the \textit{\ac{CS}} establish a secure connection over the Internet, e.g., through TLS, to register the location of the \textit{Charging Station}. Such a registration only happens once for a given \textit{Charging Station}. 
    \item[2] The \textit{Charging Station $j$} delivers its location $\mathbf{x_{CS,j}}$ to the \textit{Central Service} to store the location.
    \item[3] The \textit{Central Service} stores the location of the \textit{Charging Station} in the \ac{PvIR}, for later usage.
    \item[4] The \textit{Central Service} sends an acknowledgement message back to the \textit{Charging Station $j$}, to acknowledge correct operations, so ending the \textit{Registration Phase}.
    \item[5] At runtime, the \textit{UAV} broadcasts a \ac{RID} message through the {\em RID module}, as discussed above.
    \item[6] \textit{Charging Station 1} receives the RID message containing the UAV (obfuscated) location $\mathbf{z}$ and forwards it to the \textit{Central Service}.
    \item[7] \textit{Charging Station 2} receives the RID message containing the UAV (obfuscated) location $\mathbf{z}$ and forwards it to the \textit{Central Service}.
    \item[8] The \textit{Central Service} compares the location $\mathbf{z}$ reported by the UAV to the locations of all the \textit{Charging Stations} receiving such a message, i.e., $\mathbf{x_{CS,1}}$ and $\mathbf{x_{CS,2}}$, and determines the UID of the one which is the best to serve the request by the UAV, as $UID = i: \min_{i} (| \mathbf{z} - \mathbf{x_{CS,i}})$.
    \item[9] The \textit{Central Service} sends the UID $i$ back to the \textit{Charging Station} that is closest to the \textit{UAV}.
    \item[10] The nearest \textit{Charging Station} establish a connection with the communication module on the UAV.
    \item[11] The nearest \textit{Charging Station} sends its location to the \textit{UAV}, so that it can recharge its battery.
\end{itemize}

\begin{figure}[htbp]
\begin{center}
\includegraphics[width=8cm]{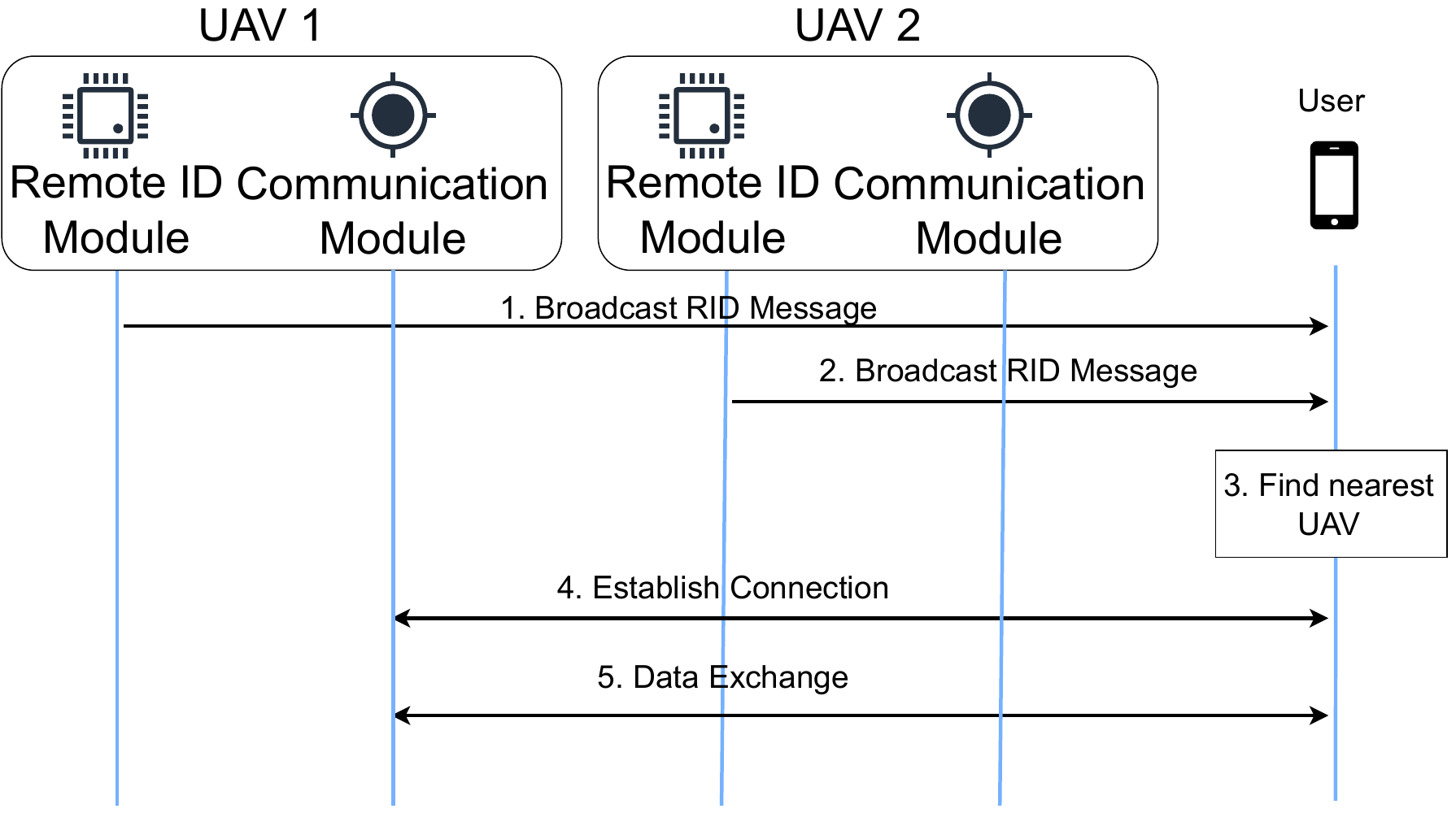}
\caption{Sequence diagram of the operations executed in Use Case \#3 to find the nearest UAV providing services.}
\label{fig:seq_service}
\end{center}
\end{figure}

\textbf{Use Case 3: UAV as a Service.} Consider a RID-enabled UAV offering telecommunications services. We show the protocol flow for this use case in Fig.~\ref{fig:seq_service} while we describe the required steps below. We omit the Registration phase since there are no changes compared to Fig.~\ref{fig:seq_complete}. 
\begin{itemize}
    \item[1] \textit{UAV 1} broadcasts a \ac{RID} message through the local {\em RID Module}, reporting its obfuscated location $\mathbf{z_1}$.
    \item[2] \textit{UAV 2} broadcasts a \ac{RID} message through the local {\em RID Module}, reporting its obfuscated location $\mathbf{z_2}$.
    \item[3] Consider a user receiving several \ac{RID} messages. It picks the nearest \textit{UAV} based on the reported obfuscated location.
    \item[4] The user and the nearest \textit{UAV}, e.g. \textit{UAV 1}, establish a secure connection .
    \item[5] The user can send and receive data to and from \textit{UAV 1}, so using its services.
\end{itemize}

\section{Security Analysis}
\label{sec:security_analysis}

In this section, we discuss the main security features offered by our solution.

{\bf Location Obfuscation Privacy.} Our solution integrates the approach proposed by Xiao et al. in~\cite{xiao2015_ccs} to disclose an obfuscated location $\mathbf{z}$ in place of the current UAV location $\mathbf{x}$. As noted by relevant works on location privacy ~\cite{chatzikokolakis2014_pets}, achieving complete location privacy when frequently disclosing the current location is not feasible. However, as demonstrated by Xiao et al. in \cite{xiao2015_ccs}, the proposed solution achieves the lower bound of trajectory privacy when considering multiple time-correlated location disclosures.\\
\indent As per the introduced modifications to the method by Xiao et al., we first notice that we apply location obfuscation on a 3-D space. As we use the algorithm independently on the horizontal (latitude and longitude) and vertical coordinates (altitude), such a change does not impact the location privacy of the single coordinates.
Our proposed \sol\ solution also modifies the computation of the prior probability array ($p_t^{-})$. Specifically, we generate all permutations of $[0,1,-1]$ and create a box around the current UAV location, consisting of $b^3$ cells, where $b$ denotes the number of cells in each dimension.
Recall that UAVs can move in space across consecutive location disclosures. Thus, such a movement may lead to the generation of new cells if the UAV moves into a new cell. In \sol, we minimize the probability of this event through the knowledge of the area where the \ac{UAV} is flying. When this change occurs, \sol\ re-initialize all cells to $p_{\#} = \frac{1}{b^3}$. As per~\cite{xiao2015_ccs}, we compute the prior probability as $p_{t}^- = p_{\#}M$, being $M$ the transition matrix.
We recall the definition of adversarial privacy~\cite{xiao2015_ccs}. 
\begin{definition}
\label{def:adversarial-privacy}
    \textbf{(Adversarial-privacy)}. A mechanism is $\epsilon$-adversarially private if for any location $\mathbf{s_i} \in \mathbf{S}$, any output $\mathbf{z}$ and any adversary knowing the true location is in $\Delta X$:
    \begin{equation}
        \frac{Pr(u^*_t = s_i|z_t)}{Pr(u_t^* = s_i)} \leq e^{\epsilon}
    \end{equation}
    where  $Pr(u_t^* = s_i)$ and $Pr(u^*_t = s_i|z_t)$ are the prior and posterior probabilities of any adversaries.
\end{definition}

In our setting, adversarial privacy is equivalent to differential privacy on the $\delta$-location, as shown by Xiao et al. \cite{xiao2015_ccs}.
Thus, we can apply the same reasoning provided by Xiao et al. in \cite{xiao2015_ccs} to prove the security of our solution, as follows.
\begin{theorem}
    \textit{At any timestamp $t$, Algorithm \ref{algo:entire_system} is $\epsilon_t$-differentially private on 0-location set}. 
\begin{proof}
    It is equivalent to proving adversarial privacy on a 0-location set, which includes all possible locations. If $x_t^* \in \Delta X_t$, then $z_t$ is generated by $x_t^*$. As shown by Xiao et al. \cite{xiao2015_ccs}, $z_t$ is $\epsilon_t$-differentially private. So $\frac{Pr(u_t^*=s_i|z_t)}{Pr(u_t^* = s_t)} \leq e^{\epsilon}$. When $x_t^* \notin \Delta X_t$, we can replace $x_t^*$ with a surrogate $\Tilde{x}_t$, and thus:
    \begin{equation}
        \resizebox{0.9\hsize}{!}{$\frac{Pr(u_t^*=s_i|z_t)}{Pr(u_t^* = s_t)} = \frac{\sum_k Pr(u_t^* = s_i| \Tilde{x}_t=s_k)Pr(\Tilde{x}_t=s_k|z_t)}{\sum_k Pr(u_t^* = s_i| \Tilde{x}_t=s_k)Pr(\Tilde{x}_t=s_k)} \leq e^{\epsilon}$}
    \end{equation}
    Therefore, being the framework adversarially private, Algorithm \ref{algo:entire_system} is also $\epsilon_t$-differentially private on 0-location set.
\end{proof}
\end{theorem}

{\bf Encrypted Location Report Robustness.} \sol\ requires delivering encrypted location reports, i.e., cryptography values allowing the TTP (USS) to obtain the plain-text \ac{UAV} location, if necessary. To do so, we use the \ac{ECIES} public-key encryption scheme, implying the generation of an ephemeral symmetric key signed through the public key of the USS. Such location reports can only be decrypted using the private key of the USS, which is assumed to stay private on the USS throughout the deployment of the system. If the private key of the USS is not leaked, based on the cryptographic properties of ECIES, a passive eavesdropper cannot obtain the plain-text location $\mathbf{x}$ used to generate the ciphertext $c$. The eavesdropper cannot even know if the plain-text location $\mathbf{x}$ changes or stays the same across consecutive location disclosures, thanks to the protection offered by ECIES against chosen-ciphertext, and chosen-plaintext attacks, as discussed in~\cite{ecies_standard}.
\section{Proof-of-Concept Implementation}
\label{sec:impl}

We implemented two \ac{PoC} of \sol, as described below.

\indent \textbf{Hardware Details.} We use two different devices as a \ac{UAV}, i.e., a general-purpose laptop and an embedded device, to evaluate the performance of \sol\ on high-end and medium-end \acp{UAV}.
Our first \ac{PoC} uses a general-purpose laptop equipped with one CPU running at 2.60GHz and 16GB of \ac{RAM}. Such features align with the resources available onboard medium-high-end \acp{UAV}, e.g., the DJI Matrice 300~\cite{djiMatrice}.
The second \ac{PoC} uses as the \ac{UAV} a \ac{RPI} Model 3B+, featuring a 1.4GHz CPU and 1GB of \ac{RAM}~\cite{raspberry_pi}. The resources of the \ac{RPI} align with the ones of medium-low-end \acp{UAV} such as the DJI Mini 2~\cite{dji_mini_2}. Also, note that such a choice aligns with the methodology used by other scientific contributions in the \ac{UAV} security domain~\cite{george2023_sacmat}. Both the laptop and the \ac{RPI} integrate an IEEE 802.11b/g/n module, useful for transmitting and receiving RID-compliant WiFi messages over the air.

\indent \textbf{Software.} We implemented \sol\ using the C programming language to reduce runtime delays deriving from language interpretation as much as possible. We used the well-known library \textit{MIRACL CORE} to implement cryptography operations required by the \ac{ECIES} encryption scheme~\cite{miracl}. Note that we chose MIRACL Core due to its well-known excellent performance on low-powered IoT systems and the available support for several elliptic curves, as required for running ECIES. Such a choice also aligns with relevant literature on security solutions for constrained \acp{UAV}~\cite{wisse2023_iotj}. We report in Tab.~\ref{tab:curve_comparison} the \acp{EC} we used within the \ac{ECIES} scheme. To show the difference between curves of the same security levels, we have also considered two additional curves, namely BN254 and BLS48556. Using such curves could give us information about the difference in performance between two curves with the same level of security. Note that the curve BN254 was designed for 128-bit security \cite{bn254} but, recently, its actual security decreased to 100-bit~\cite{bn254_issues}.
\begin{table}[htbp]
    \centering
    \caption{ECC Curves features.}
    \label{tab:curve_comparison}
    \begin{tabular}{c|c|c}
    {\bf ECC Curve} & {\bf Key size [B]}    & {\bf \#-Bit Security} \\ \hline
    BN254  & 64 &  100~\cite{bn254_issues}\\
    secp256r1 (NIST256)  &  64 & 128~\cite{nist_curves}\\
    secp384r1 (NIST384)  & 96 & 192~\cite{nist_curves}\\
    secp521r1 (NIST521)  & 132 &  256~\cite{nist_curves}\\
    BLS48556 &  140 &  256~\cite{bls48} \\
    \end{tabular}
\end{table}

As for the extended \ac{PIM} solution integrated into \sol, we first implemented the required operations using MATLAB\textcopyright\ R2023a. Then, we used the tool \textit{MATLAB Coder} to convert the code into C. We manually fixed a few post-conversion inconsistencies and then cross-compiled the code through \textit{GCC} and \textit{GNU make} to create a library able to run on the target hardware devices used in our PoC.
Tab.~\ref{tab:message_format} shows the message format of \sol. 
\begin{table}[htbp]
\centering
\caption{Modified RID message format in Bytes.}
\label{tab:message_format}
\begin{tabular}{c|c}
{\bf Content}            & {\bf Bytes [B]}   \\ \hline
Unique identifier        & 4       \\
Obfuscated UAV longitude & 4       \\
Obfuscated UAV latitude  & 4       \\
Obfuscated UAV altitude  & 4       \\
UAV longitude velocity   & 2       \\
UAV latitude velocity    & 2       \\
UAV altitude velocity    & 2       \\
CS longitude             & 4       \\
CS latitude              & 4       \\
CS altitude              & 4       \\
Timestamp                & 4       \\
Emergency status         & 1       \\
Ephemeral key            & 64-140  \\
Ciphertext               & 16      \\
HMAC tag                 & 32      \\ \hline
Total                    & 151-227
\end{tabular}
\end{table}

We ran \sol\ on the \ac{RPI} as a native process. On the laptop, to isolate it from concurrent processes, we ran \sol\ within a Linux-based virtual environment featuring 4 CPU cores and 8GB of \ac{RAM}. The source code of our solution is available open-source at~\cite{code}.

\section{Experimental Evaluation}
\label{sec:evalution_and_results}


\subsection{Experimental Methodology} \label{subsec:research_questions_and_reference_dataset}

{\bf Research Questions.} Our experimental assessment aims to answer the following \acp{RQ}.
\begin{itemize}
    \item \textbf{RQ1:} What is the runtime overhead of \sol\ on regular \acp{UAV} regarding computation time, \ac{RAM} usage, energy expenditure, and communication bandwidth with various security levels?
    \item \textbf{RQ2:} How can we improve the location privacy provided to the \ac{UAV} at runtime, and how do such improvements affect the overhead of our solution?
    \item \textbf{RQ3:} What is the utility of the location disclosed through \sol\ by the \ac{UAV} across various configuration parameters while considering our reference use cases discussed in Sec.~\ref{subsec:use_cases}?
\end{itemize}

\acp{UAV} are constrained devices, characterized by (one or more) constraints regarding available computational, memory, communication bandwidth, and energy resources. Such constraints make it challenging to implement and run a suitable technique for location obfuscation while still creating and broadcasting \ac{RID} messages within the time limitations imposed by the RID regulation ($1$~second). 
\textbf{RQ1} sets a baseline for the overhead of our solution when using different hardware and various \acp{EC}. \textbf{RQ2} investigates the impact in terms of location privacy and overhead on the \ac{UAV} derived by changing relevant location obfuscation parameters (i.e., $\epsilon$).  
Finally, \textbf{RQ3} evaluates \sol\ in the specific context of our use cases (Sec.~\ref{subsec:use_cases}). We specifically focus on the trade-off between \ac{UAV} location privacy and utility, i.e., the extent to which \ac{UAV}-based \acp{LBS} can successfully use the disclosed location.

{\bf Experiments Settings.} To answer the RQs above, we designed a measurement system to obtain the execution time, \ac{RAM} usage, communication overhead, energy consumption, location privacy and utility of our approach.
As for the execution time, we measured the time (in milliseconds) necessary to generate a \ac{RID} message secured through \sol, from the acquisition of the current location from the GNSS (not included) to the assembling of the entire RID message (wireless communication not included). We specifically excluded location acquisition and wireless communication as they are orthogonal processes to our solution, which might be executed asynchronously. As for the RAM usage, we used the Linux tool \emph{/usr/bin/time -v}, providing the maximum instantaneous RAM occupancy of a process. Through this tool, we can obtain an indication of the amount of RAM to be deployed on the involved devices to support our solution fully. As for the communication cost, using standard tools provided by default C libraries, we measured the size of the complete RID message after the application of our solution. We also measure the energy consumption (in milliJoule) the \ac{UAV} requires to run our solution. To this end, in line with similar contributions, we used the energy estimation tool \emph{powertop} provided by the Linux OS. Powertop estimates the power (in Watts) of a running process based on hardware-dependent power consumption figures. To compute the corresponding energy consumption $E$, we multiply the obtained power estimate $P$ for the duration of the process of interest $\Delta T$ (the duration of the runtime phase), measured as described above, as $E = P \times \Delta T$. Such a value can indicate the energy required to run our solution and, thus, the battery to be installed on the \ac{UAV} to keep broadcasting (location) privacy-preserving RID messages for the whole mission. 
To evaluate the location privacy provided to the \ac{UAV} thanks to the application of our solution, in line with the relevant literature~\cite{brighente2022_ares}, we use the \emph{average distance}, i.e., the average value of the (absolute) difference between the actual \ac{UAV} location and the disclosed (obfuscated) one. The higher the average distance, the more significant the uncertainty on the real \ac{UAV}'s location, thus, the higher the location privacy. 
Finally, as for the utility of the \ac{UAV}'s disclosed location, we ran simulations in Matlab using real \ac{UAV} traces (see below the reference dataset used in our analysis). We used metrics specific to the considered use case. For the use case \#1, we considered the number of \acp{TP} (actual detected invasions of no-fly area), \acp{FP} (false detected invasions of no-fly area), \acp{TN} (correct identified non-invasions of the no-fly area) and \acp{FN} (missed detected invasions of the no-fly area) as seen by the invasion monitoring system. For the use case \#2, we considered the \emph{average extra distance}, i.e., the extra distance the \ac{UAV} has to travel to reach the suggested charging station compared to the distance to reach the optimal one. Finally, for the use case \#3, we evaluate the \emph{average extra distance} as the additional space the signal emitted from the suggested UAV has to travel to reach the user compared to the distance it would travel from the UAV closest to the user. 

\begin{figure}[htbp]
    \begin{center}
    \includegraphics[width=.7\columnwidth]{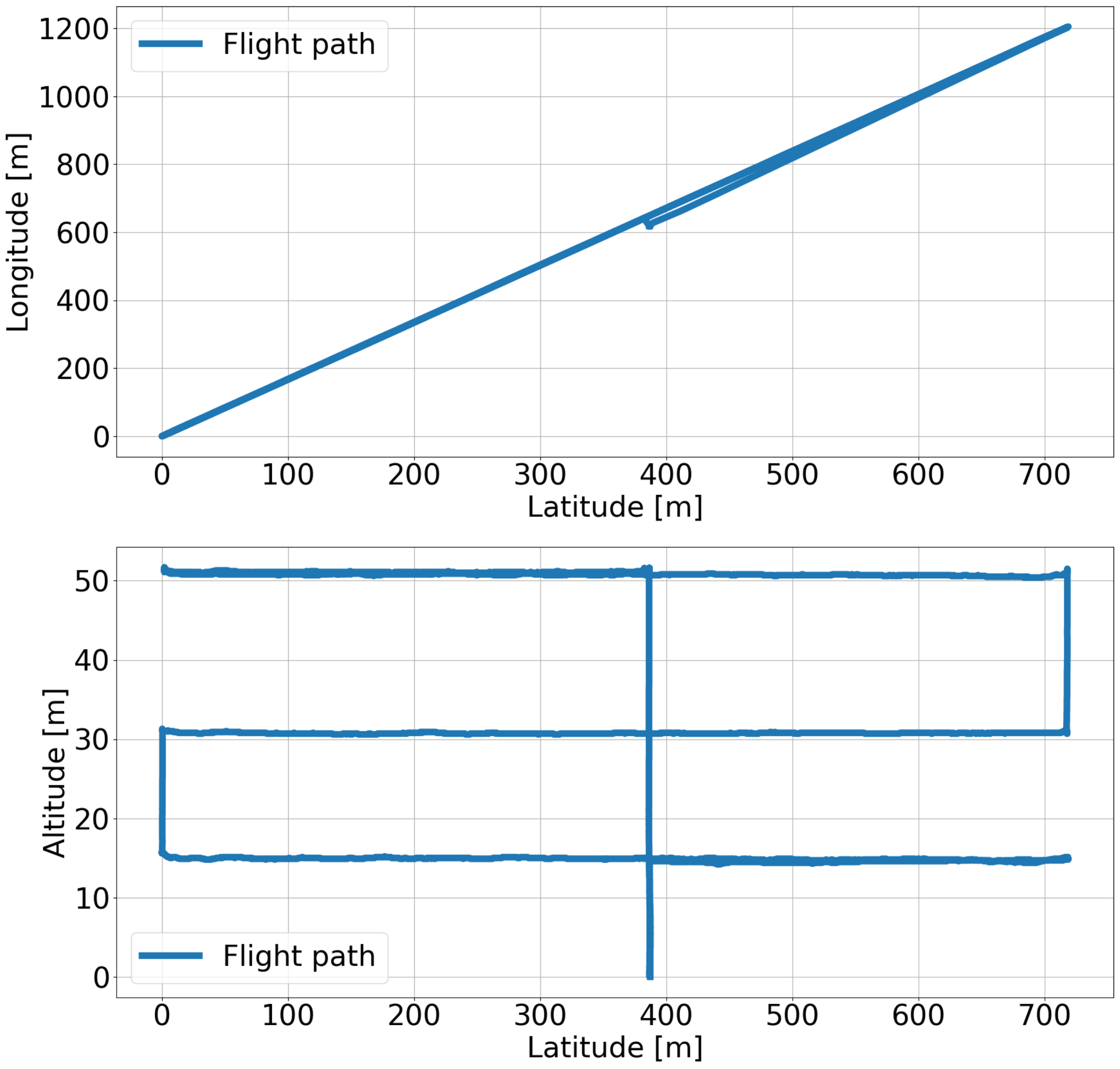}
    \caption{Representation of the flight path used for the experiments in Sec.~\ref{subsec:experiment_1_overhead_baseline} and \ref{subsec:experiment_2_epsilon_delta}, with top-down and side view.}
    \label{fig:flight_path}
    \end{center}
\end{figure}

\begin{figure}[htbp]
    \begin{center}
    \includegraphics[width=.7\columnwidth]{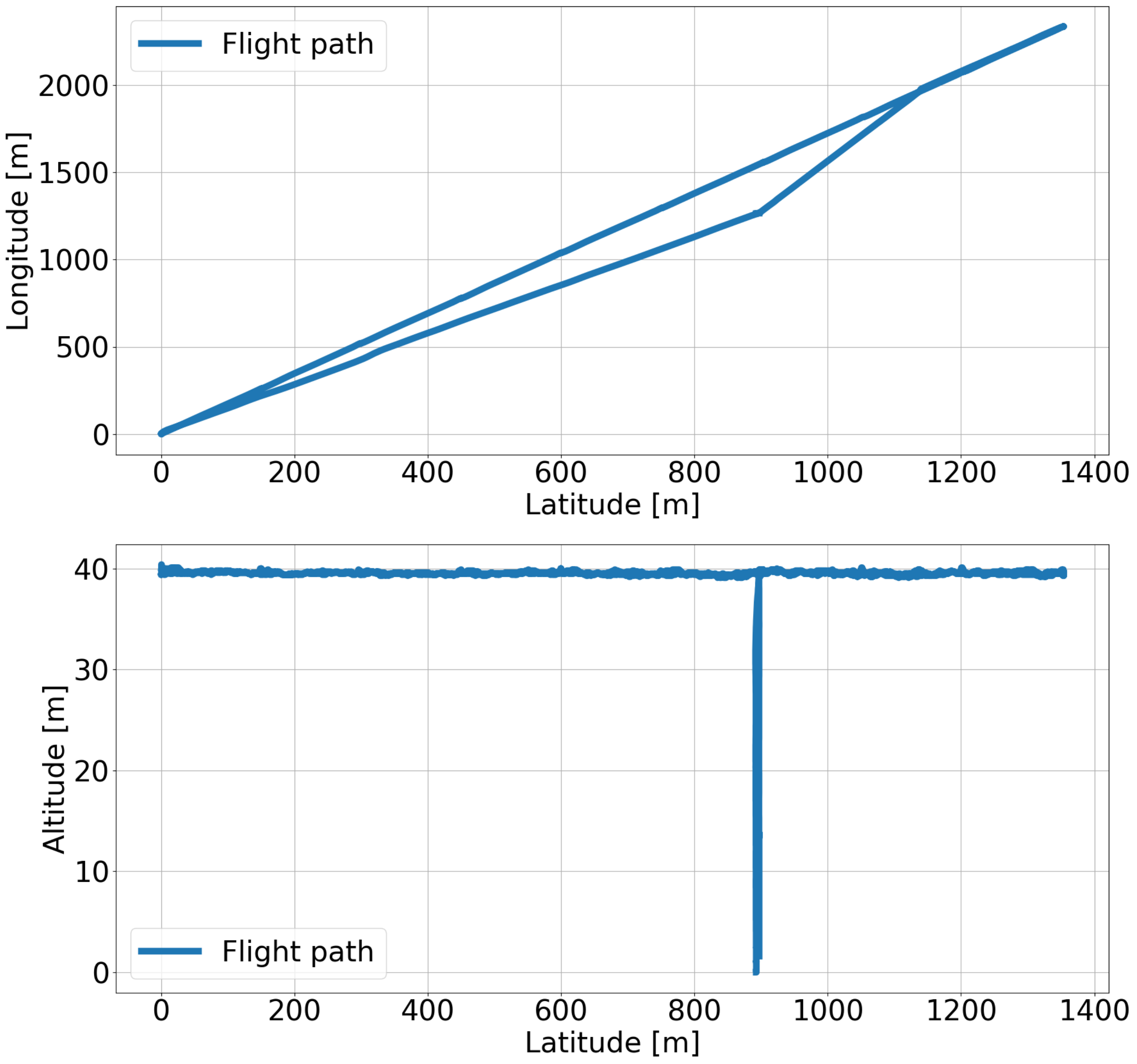}
    \caption{Representation of the flight path used for the experiments in Sec.~\ref{subsec:use_case_experimentations}, with top-down and side view. }
    \label{fig:flight_path_usecases}
    \end{center}
\end{figure}

\begin{figure}[htbp]
    \begin{center}
    \includegraphics[width=.7\columnwidth]{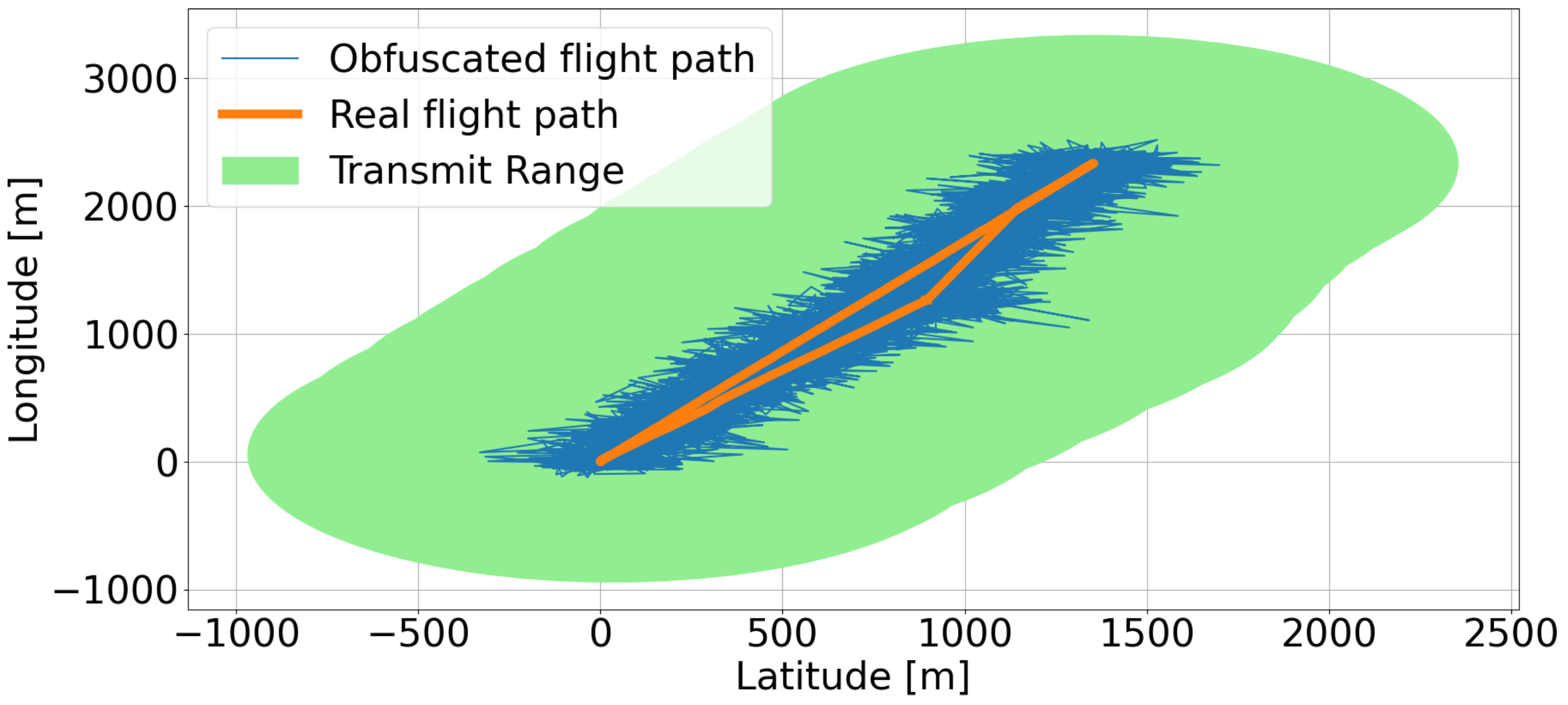}
    \caption{UAV actual trajectory, obfuscated trajectory (with an average distance of 100 meters) and WiFi coverage of $1,000$~meters.}
    \label{fig:obfuscated}
    \end{center}
\end{figure}

\textbf{Reference Dataset.} In this work, for flight data of the involved \acp{UAV}, we use the actual flight dataset provided by NATO's Emerging Security Challenges Division~\cite{flight_data}, in line with similar works~\cite{brighente2022_ares}. The data refer to multiple flights paths of various \acp{UAV} in an area of $1.5$ km $\times$ $2.6$ km $\times$ $40$~m. For all the tests, we use the flight path shown in Fig.~\ref{fig:flight_path}, including over $5,000$ location disclosures in its flight path. Instead, for the analysis of the use cases, we use the flight path shown in Fig.~\ref{fig:flight_path_usecases}, which includes over $8,000$ locations. We used a different flight path to analyze the use cases so as to work in a controlled environment, where the UAV flies mainly in a straight line. Such a trajectory eliminates anomalies due to weird flight paths and makes it easier to reason about the entire scenario.
We ran all tests at least $5,000$ times, so as to average the results over multiple locations and trajectories. We also ran all tests using the highest privileges on the specific hardware, avoiding delays due to background concurrent processes. As a reference, we show in Fig.~\ref{fig:obfuscated} the relationship between the actual \ac{UAV} path, the released obfuscated locations and the WiFi coverage (upper bound on the location privacy).

\subsection{Experiment 1: Overhead Evaluation}
\label{subsec:experiment_1_overhead_baseline}

\begin{figure}[htbp]
    \begin{center}
    \includegraphics[width=.7\columnwidth]{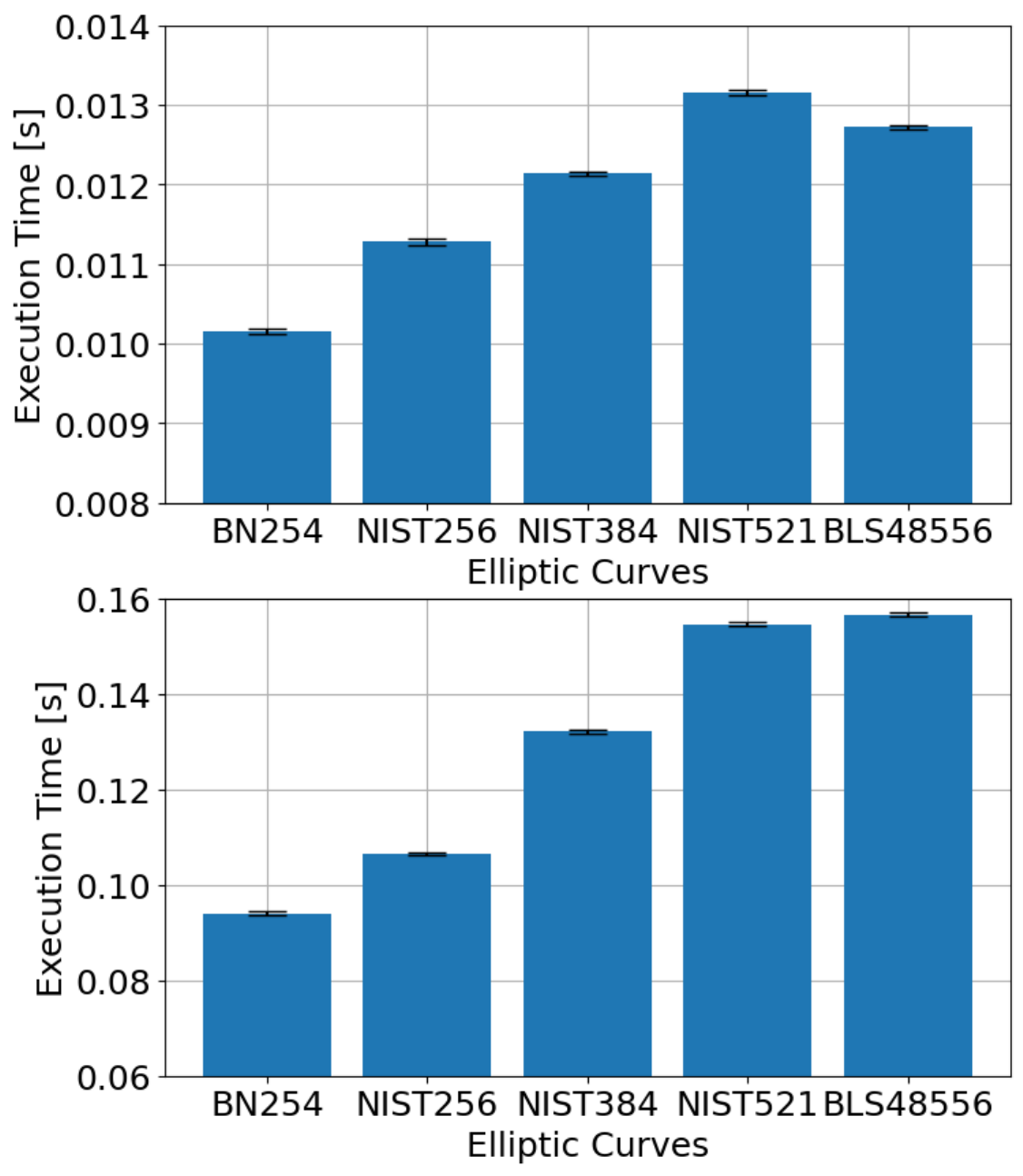}
    \caption{Time required to run \sol\ on a regular-laptop (up) and \ac{RPI} (bottom), with 95\% confidence interval.}
    \label{fig:basic_time}
    \end{center}
\end{figure}

To answer \textbf{RQ1}, we measure the overhead of \sol\ with different curves on a regular laptop and a \ac{RPI}.\\
\indent \textbf{Time.} Fig.~\ref{fig:basic_time} reports the average time (and 95\% confidence intervals) required to create a RID message using \sol. The upper figure shows the results of our tests on the laptop, while the bottom results are for the \ac{RPI}.

As intuition would suggest, the average time required to run our solution increases with the size of the EC points (keys) used in the curve, i.e., with the provided security level. Considering the two curves NIST521 and BLS48556, both providing 256-bit security, we can see a slight discrepancy between the results obtained on the laptop and those obtained on the \ac{RPI}. However, this difference is negligible ($3.5$~\%), and it does not affect the rationale of our findings. It is essential to notice that all our results, both on the laptop and on the \ac{RPI}, report execution times well below the threshold of $1$~s. The minimal execution times demonstrate the viability of our solution on relevant hardware.

\begin{figure}[htbp]
    \begin{center}
    \includegraphics[width=.7\columnwidth]{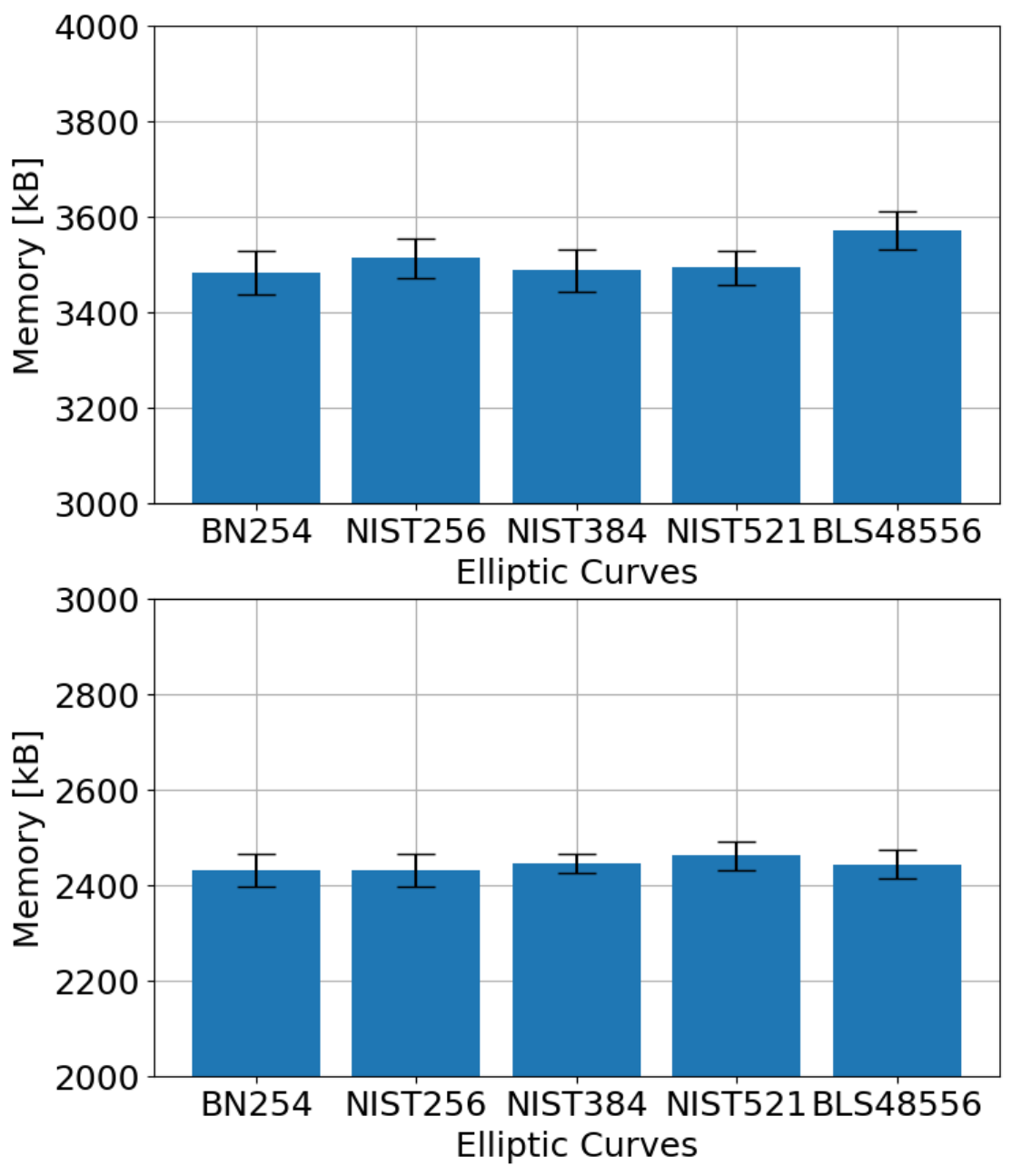}
    \caption{RAM required by \sol\ on the general-purpose laptop (upper figure) and \ac{RPI} (bottom), with 95\% confidence interval.}
    \label{fig:basic_memory}
    \end{center}
\end{figure}

\indent \textbf{RAM.} Fig.~\ref{fig:basic_memory} reports the average peak RAM consumption of our solution, with 95\% confidence intervals.
We notice that, on the particular hardware platform, the RAM consumption of our solution is independent of the chosen security level. On the laptop, our solution requires $3,600$~kB of RAM in the worst case, which is very lightweight and affordable even on low-end \acp{UAV}. We highlight that it is not possible to cross-compare the RAM consumption on the two different platforms because the underlying memory management system of such platforms is different. 

\indent \textbf{Communication.} To evaluate the communication overhead of \sol, we can compare the regular RID format with the RID message format required by our solution, summarized in Table~\ref{tab:message_format}. \sol\ requires changing the content of three fields (current latitude, longitude, and altitude of the \ac{UAV}) and delivering three additional fields. The modified fields require the same amount of bytes as the regular fields, generating no additional overhead. 
As for the three added components, i.e., ephemeral key, ciphertext, and HMAC tag, the ciphertext and HMAC tag required by ECIES have a constant size of 16 and 32 bytes, respectively. The size of the ephemeral key depends on the adopted curve and, as per Tab.~\ref{tab:curve_comparison}), it varies between 64 and 140 bytes.
Thus, the total communication overhead of \sol\ for a single message ranges between  $16+32+64=112$~Bytes and $16+32+140=188$~Bytes. The maximum RID message size with \sol\ amounts to $227$~bytes, which is well below WiFi's \ac{MTU}, i.e., 2,304 bytes~\cite{ieee2012}.

\begin{figure}[htbp]
    \begin{center}
    \includegraphics[width=.7\columnwidth]{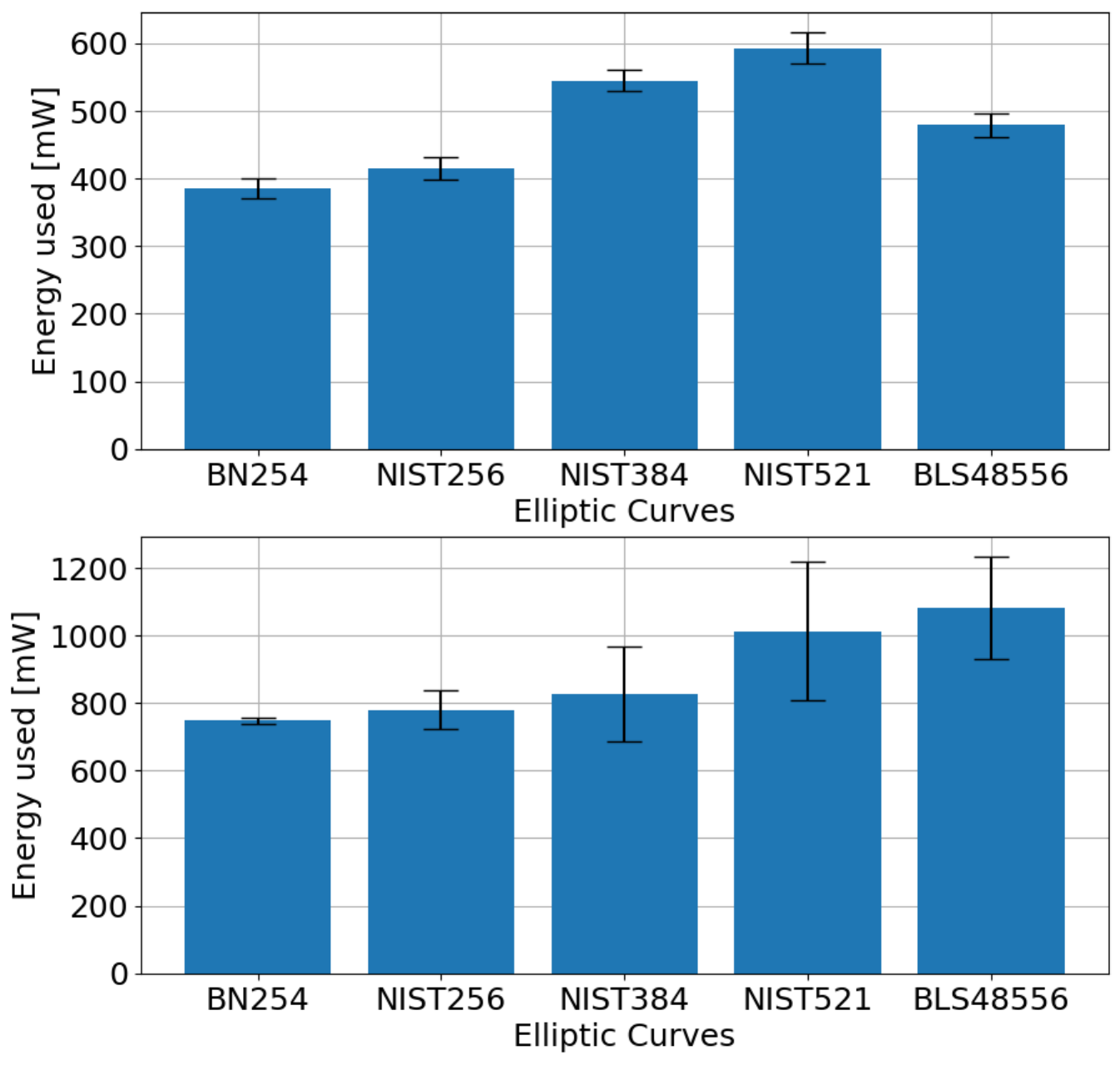}
    \caption{Energy expenditure of \sol\ on a regular laptop (upper figure) and Raspberry PI 3 (bottom), with 95\% confidence interval.}
    \label{fig:basic_energy}
    \end{center}
\end{figure}

\indent \textbf{Energy.} Fig.~\ref{fig:basic_energy} reports the average energy required by \sol\ on both the laptop and the \ac{RPI}, together with 95\% confidence intervals.
In line with Fig.~\ref{fig:basic_time}, increasing the number of security bits increases the energy consumption of our solution. The energy expenditure is primarily influenced by the time necessary to complete the protocol. We can notice the same difference between the curves NIST521 and BLS48556 among the two testbeds, as discussed above. The most energy-consuming curve on the laptop is NIST521, which requires $600$~mJ for protocol run. The most energy-consuming curve on the \ac{RPI} is BLS548556, which requires an average of $1,100$~mJ per protocol run. We cannot cross-compare the energy expenditure across different hardware platforms, due to the different components and voltage needed to power them. We discuss the impact of the measured energy expenditure on an average \ac{UAV} in Sec.~\ref{sec:discussion}.

\subsection{Experiment 2: Privacy Analysis}
\label{subsec:experiment_2_epsilon_delta}

\begin{figure}[htbp]
    \begin{center}
    \includegraphics[width=.8\columnwidth]{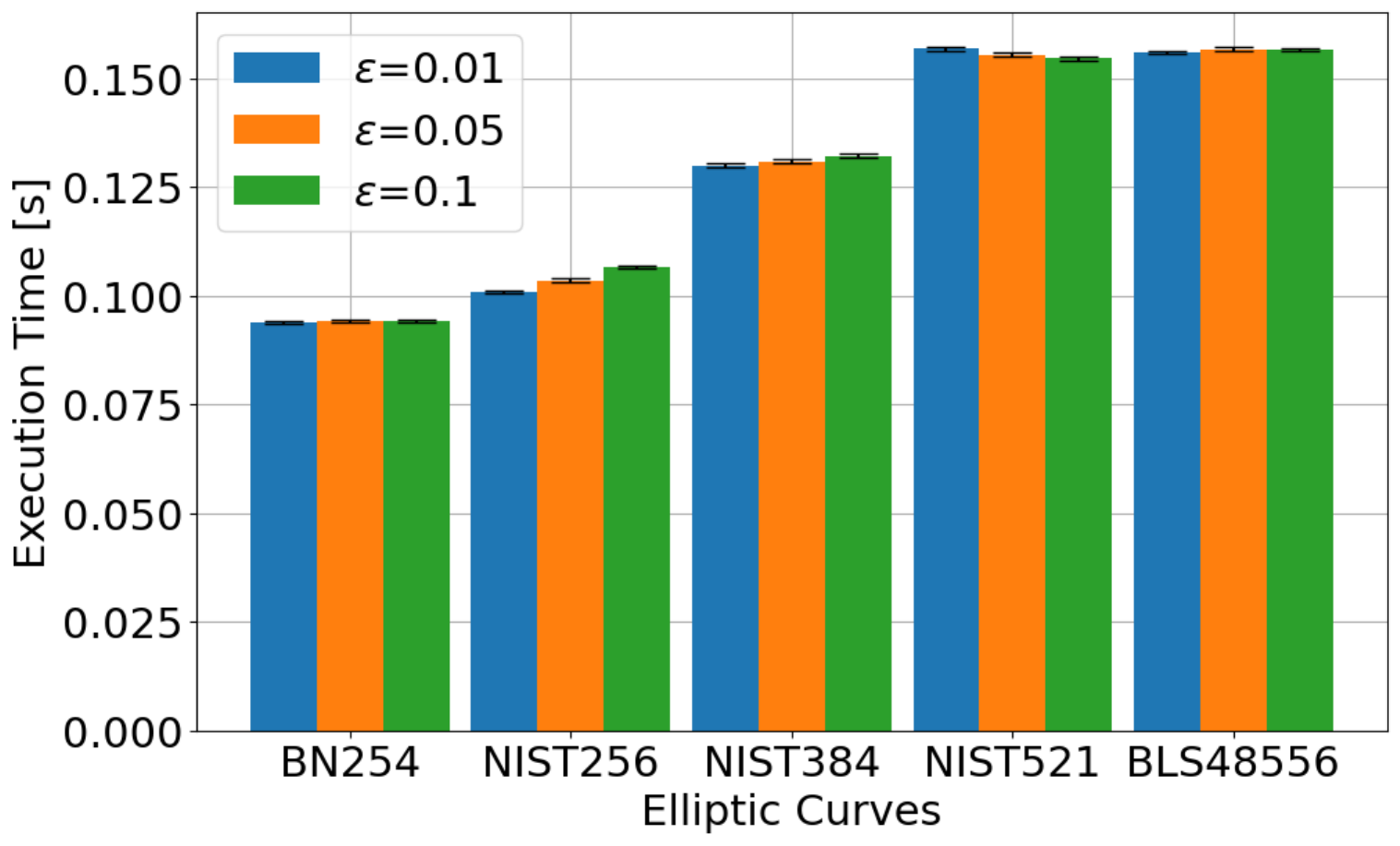}
    \caption{Time required to compute a RID message with \sol\ and various values of $\epsilon$ on a \ac{RPI}, with 95\% confidence interval.}
    \label{fig:eps_time}
    \end{center}
\end{figure}

\begin{figure}[htbp]
    \begin{center}
    \includegraphics[width=.8\columnwidth]{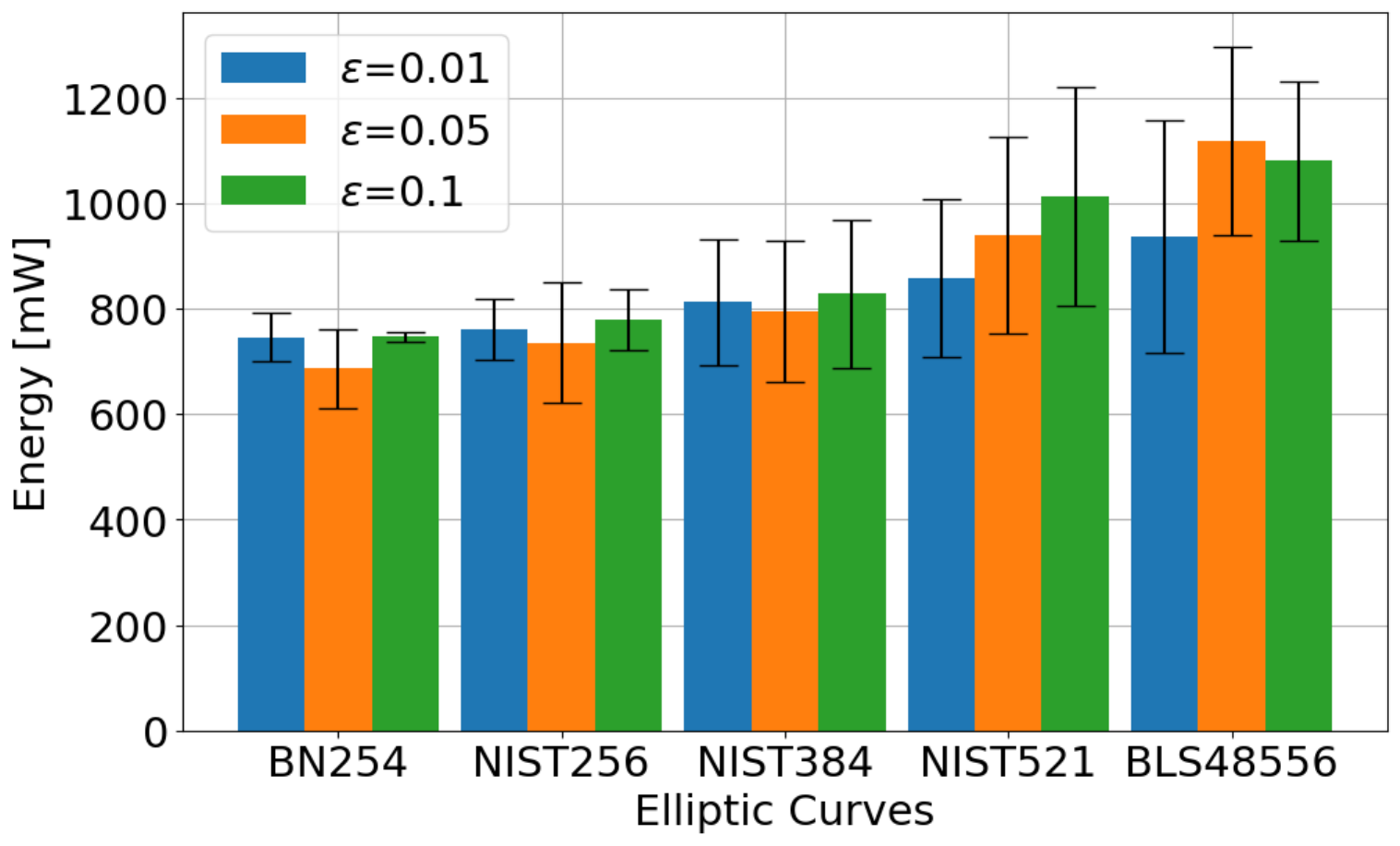}
    \caption{Energy required to compute a RID message with \sol\ and various values of $\epsilon$ on a \ac{RPI}, with 95\% confidence interval.}
    \label{fig:eps_energy}
    \end{center}
\end{figure}

\textcolor{black}{For answering \textbf{RQ2}, we analyze the impact of $\epsilon$ (privacy budget) on the location privacy provided to the \ac{UAV} and the overhead incurred by \sol. We recall from Sec.~\ref{sec:core_proposal} that \sol\ uses $\epsilon$ to tune the privacy level of the extended \ac{PIM} obfuscation mechanism.
Tab.~\ref{tab:epsilon_comparison} reports the average distance of the disclosed location to the actual one for different values of the parameter $\epsilon$, with $\delta=0.01$.}
\begin{table}[!t]
\centering
\caption{Impact of the parameter $\epsilon$ on the average distance between the actual UAV location and the disclosed (obfuscated) one, with $\delta=0.01$.}
\label{tab:epsilon_comparison}
\begin{tabular}{c|c}
$\epsilon$ & Average Distance [m] \\ \hline
0.001                   & 2533.38          \\
0.005                   & 506.65           \\
0.01                    & 253.37           \\
0.05                    & 50.72            \\
0.1                     & 25.35            \\
0.5                     & 5.06             \\
1                       & 2.54            \\
\end{tabular}
\end{table}
\indent \textcolor{black}{Changing $\epsilon$ affects the average distance significantly, from just a few meters ($2.54$~meters with $\epsilon=1$) to a few kilometres ($2,533.38$~meters with $\epsilon=0.001$). On the \ac{UAV}, lower values of $\epsilon$ provide higher average distance values and, thus, more privacy.} However, we have to consider also the utility of the disclosed location to \acp{LBS}. We analyze the utility of the disclosed locations in Sec.~\ref{subsec:use_case_experimentations} while, in the rest of this analysis, we focus on three reference values of $\epsilon$, i.e., $0.01, 0.05, 0.1$.
Fig.~\ref{fig:eps_time} and Fig.~\ref{fig:eps_energy} report the average time and the energy (with 95\% confidence intervals) required to compute a RID message through our solution on the \ac{RPI} with such values of $\epsilon$, respectively.

\begin{figure}[htbp]
    \begin{center}
    \includegraphics[width=.8\columnwidth]{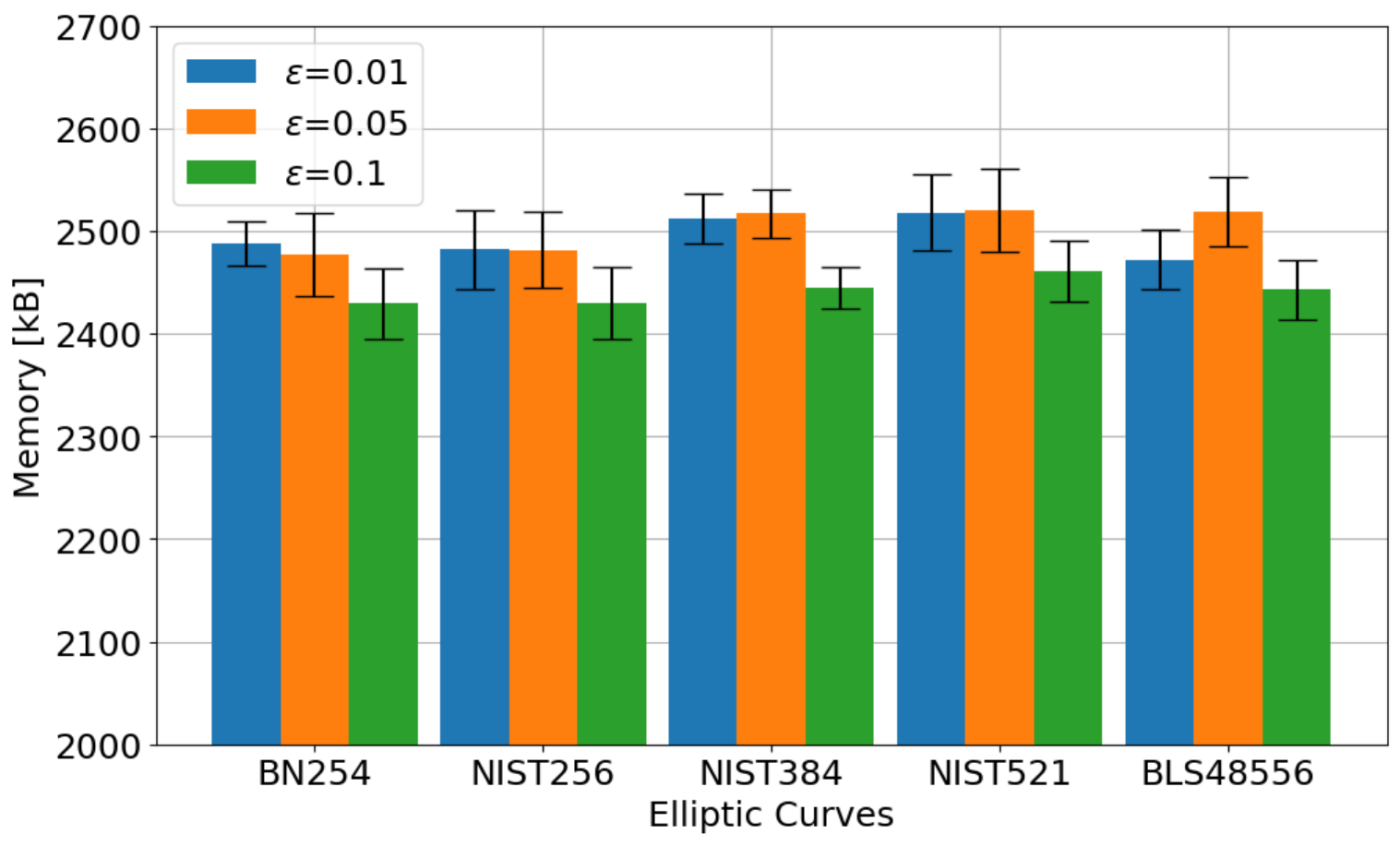}
    \caption{Memory required to compute a RID message with \sol\ and various values of $\epsilon$ on a \ac{RPI}, with 95\% confidence interval.}
    \label{fig:eps_memory}
    \end{center}
\end{figure}

From Fig.~\ref{fig:eps_time}, we can see no impact on the execution time when we change $\epsilon$. We notice minor changes among the different $\epsilon$ values. In general, $\epsilon=0.1$ uses more energy than the others---in the worst case, $18$\% more. When looking specifically at the higher security \acp{EC} $\epsilon=0.01$ is more energy efficient. Overall, the trend found in Fig.~\ref{fig:basic_time} still applies, i.e., the EC curve used as part of ECIES is the most relevant source of energy consumption in the protocol.
We also notice oscillating results when investigating the protocol's RAM consumption while varying $\epsilon$, as shown in Fig.~\ref{fig:eps_memory}. Here, we notice that choosing $\epsilon=0.1$ is always more memory-efficient than the other options, but with a small margin of $3$\%, while $\epsilon=0.01$ and $\epsilon=0.05$ approximately use the same amount of \ac{RAM}. As all of the values of $ \epsilon $ report similar results, we believe the specific value of $ \epsilon $ does not affect the RAM overhead of our solution significantly.




\subsection{Experiment 3: Use Cases} 
\label{subsec:use_case_experimentations} 

To answer {\bf RQ3}, we consider each use case introduced and described in Sec.~\ref{subsec:use_cases}. For each of them, we identify utility criteria and investigate the impact of different protocol configurations on the utility of the disclosed location.

\begin{figure}[htbp]
    \begin{center}
    \includegraphics[width=.7\columnwidth]{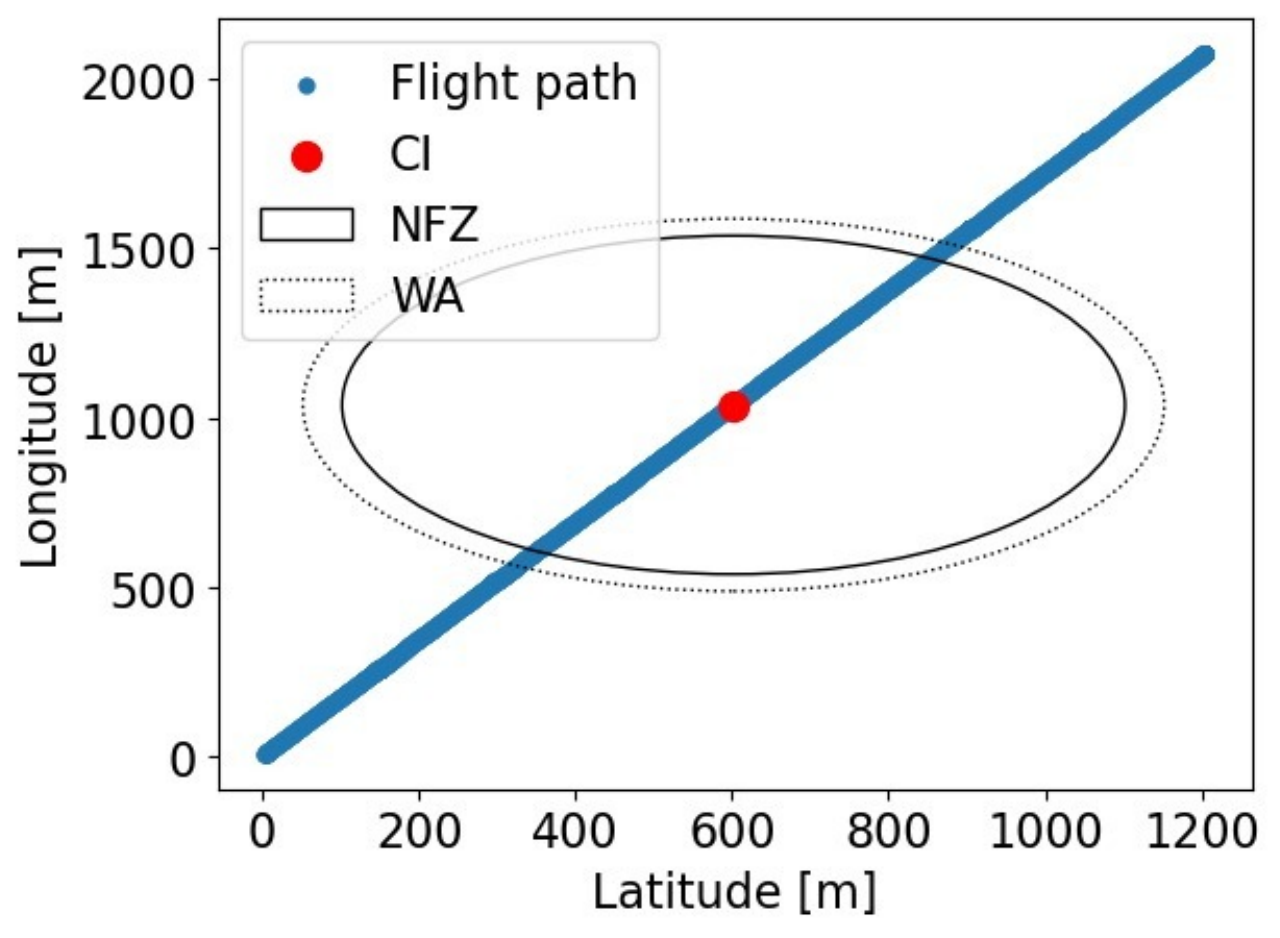}
    \caption{Use Case \# 1 analysis: sketch of the \ac{NFZ} and \ac{WA} surrounding a \ac{CI}, with an actual \ac{UAV} flight path passing through.}
    \label{fig:nfz_sketch}
    \end{center}
\end{figure}

\begin{figure}[htbp]
    \begin{center}
    \includegraphics[width=.7\columnwidth]{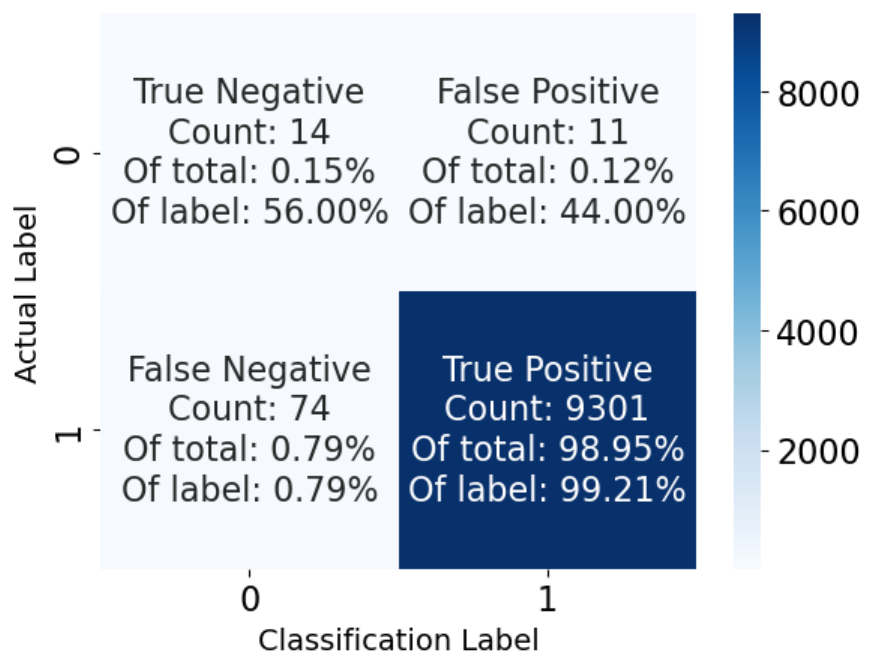}
    \caption{Confusion matrix of the detection system in Use Case \#1, considering a \ac{WA} radius of 505~m and an average distance of 25~m.}
    \label{fig:cm_r_5_avg_25}
    \end{center}
\end{figure}

\begin{figure}[htbp]
    \begin{center}
    \includegraphics[width=.7\columnwidth]{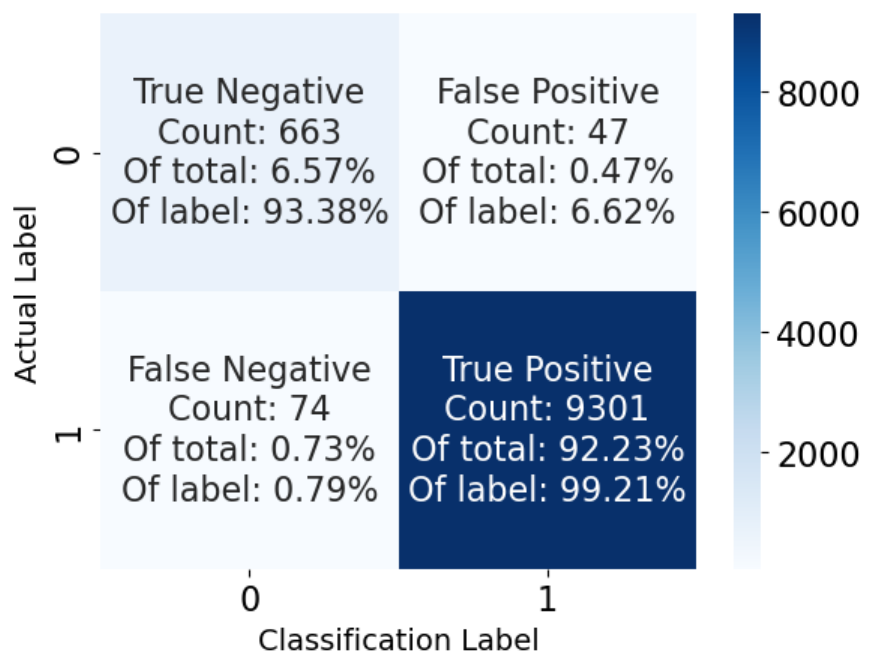}
    \caption{Confusion matrix of the detection system in Use Case \#1, considering a \ac{WA} radius of 600~m and an average distance of 25~m.}
    \label{fig:cm_r_100_avg_25}
    \end{center}
\end{figure}

\begin{figure}[htbp]
\begin{center}
\includegraphics[width=.7\columnwidth]{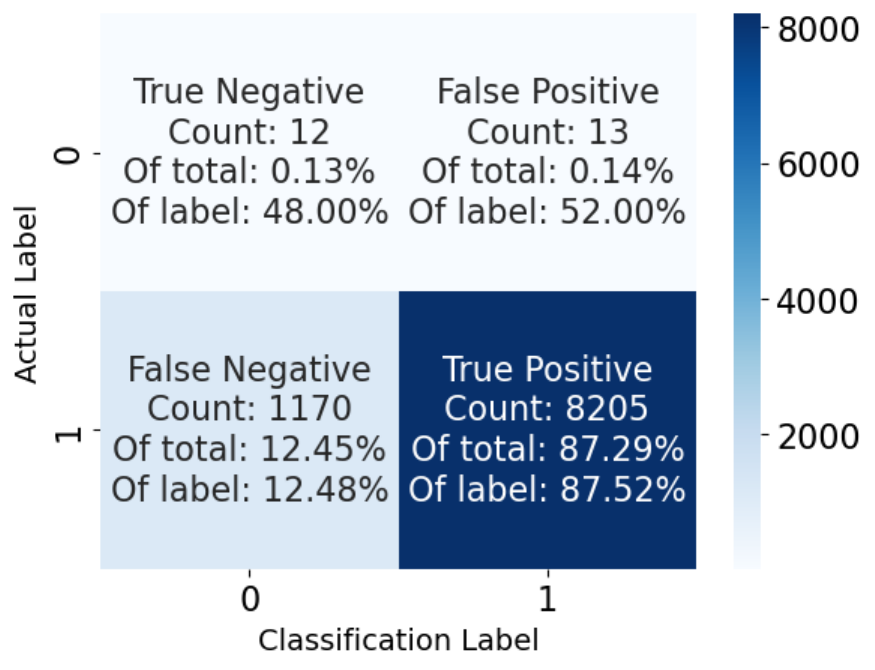}
\caption{Confusion matrix of the detection system in Use Case \#1, considering a \ac{WA} of 505~m radius and an average distance of 100~m.}
\label{fig:cm_r_5_avg_100}
\end{center}
\end{figure}

\begin{figure}[htbp]
\begin{center}
\includegraphics[width=.7\columnwidth]{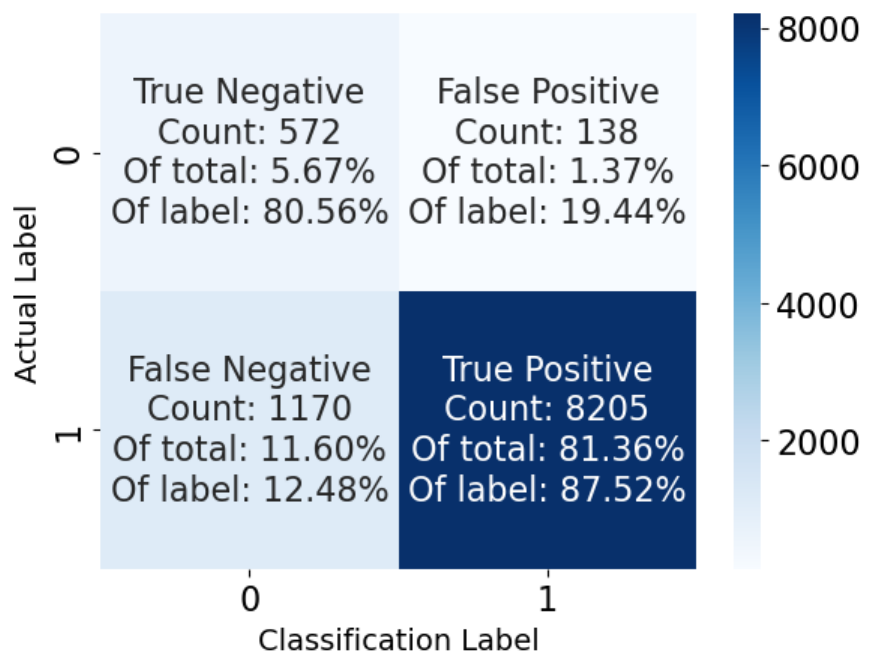}
\caption{Confusion matrix of the detection system in Use Case \#1, considering a \ac{WA} of 600~m radius and an average distance of 100~m.}
\label{fig:cm_r_100_avg_100}
\end{center}
\end{figure}

\indent \textbf{Use Case 1: Invasion of No-Fly Zone.} Consider the Use Case 1 described in Sec.~\ref{subsec:use_cases} and particularly, the scenario depicted in Fig.~\ref{fig:nfz_sketch}, where the big red circle represents the CI location, the straight black path represents the NFZ boundaries, the dotted black path represents the \ac{WA} bounds, and the blue path identifies the \ac{UAV} path.
Recall the definitions of \acp{TP}, \acp{FP}, \acp{TN}, and \acp{FN} given in Sec.~\ref{subsec:research_questions_and_reference_dataset}. In general, we aim to have very high \acp{TP} and \acp{TN}, as well as very low \acp{FP} and \acp{FN}. However, as timely \ac{UAV} detection takes priority, we aim to keep the \acp{TP} high while reducing the \acp{FP} as much as possible. 
We consider two reference values for the average distance, i.e., 25~m and 100~m. We set the radius of the NFZ to 500~m, and we consider two values for the radius of the WA, i.e., 505~m and 600~m, to investigate to what extent the difference between the NFZ and the WA impacts our findings.
Figs.~\ref{fig:cm_r_5_avg_25} and \ref{fig:cm_r_100_avg_25} report the confusion matrices of the detection system considering the average distance of 25~m and the \ac{WA} radius of 505 and 600~m, respectively.
We notice that the size of the WA does not impact the \acp{TP}, while it does affect the \acp{FP}. In fact, increasing the radius of the WA, even though the percentage of \acp{FP} decreases, the actual number of \acp{FP} increases. Considering that each \ac{FP} involves an interaction over the Internet between the CI observer and the USS (see Sec.~\ref{sec:proto_usecases}), the CI operator can choose a small \ac{WA}, characterized by a radius similar to the one of the \ac{NFZ}, so as to reduce \acp{FP} and, accordingly, the required interactions with the USS. We can notice the trend described above also when comparing Figs.~\ref{fig:cm_r_5_avg_100} and \ref{fig:cm_r_100_avg_100}, considering the same analysis as above but with an average distance of $100$~m in the locations disclosed by the \ac{UAV}. Increasing the average distance to $100$~m increases the \acp{FN} significantly compared to using an average distance of $50$~meters, thus increasing the chances that the CI misses an invading \ac{UAV}. At the same time, with an average distance of $100$~m, the \acp{FP} also increase, as there are more chances that a \ac{UAV} which is not invading discloses a location inside the NFZ, thus falsely resulting as invading. 
To decrease the number of \acp{FP}, the \ac{WA} should be as small as possible. While increasing the average distance increases \acp{FN}, there are more chances that the invasion is detected within a few seconds, resulting in a \ac{TP}. Also, when the \ac{WA} is small, increasing the average distance barely affects the number of \acp{FP}.

\begin{figure}[htbp]
\begin{center}
\includegraphics[width=.6\columnwidth]{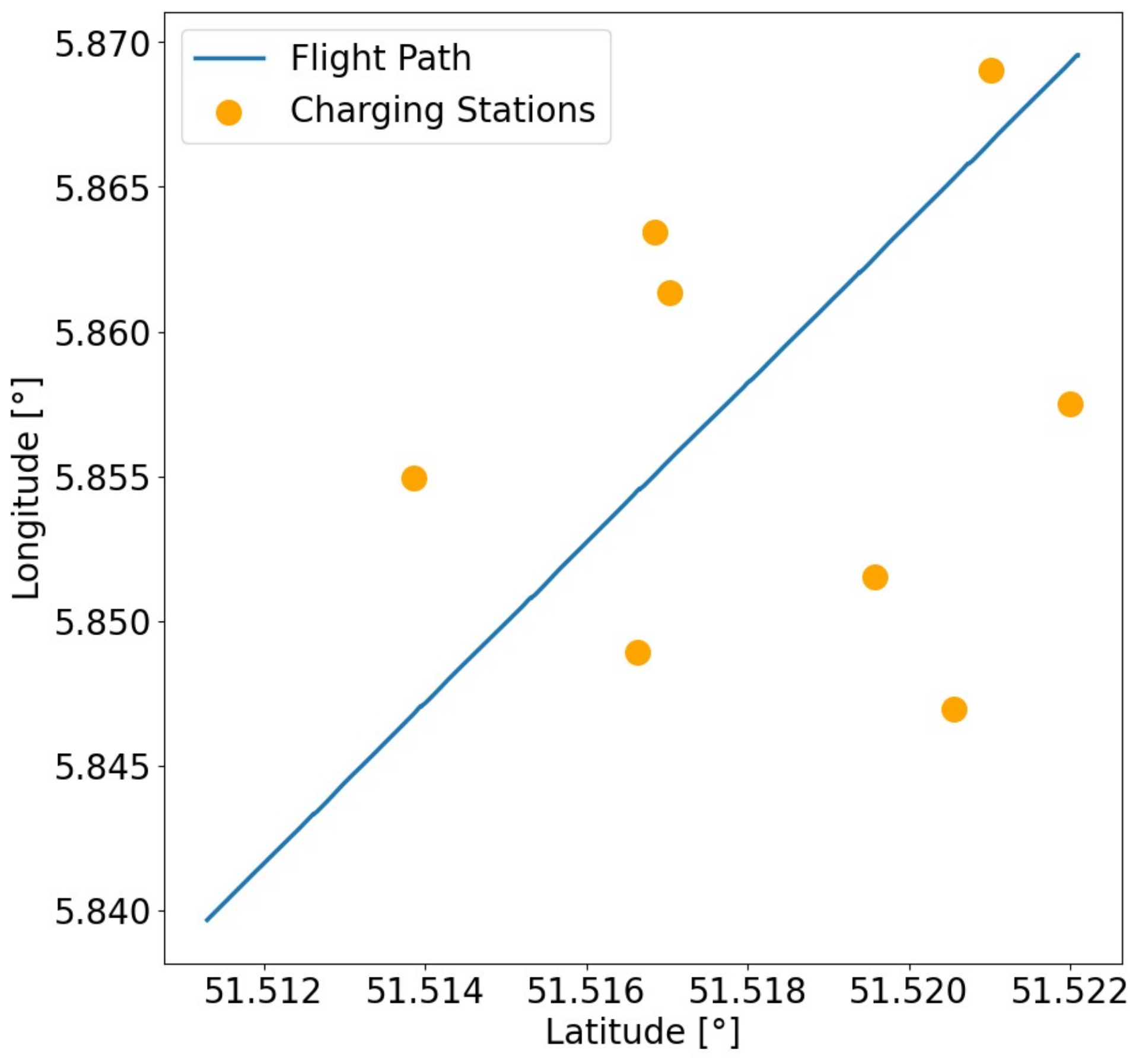}
\caption{Use Case \# 2 analysis: a sketch of a UAV flight path with eight charging stations.}
\label{fig:exp_charging_stations}
\end{center}
\end{figure}

\indent \textbf{Use Case 2: Discovery of the closest Charging Station.} Consider the Use Case 2 described in Sec.~\ref{subsec:use_cases}, and particularly, the scenario depicted in Fig.~\ref{fig:exp_charging_stations}, where an UAV is moving in area where $8$ charging stations are deployed.

Without \sol, the system would suggest to the UAV the nearest charging station to the actual UAV location. With \sol\ in place, as the UAV discloses an obfuscated location, the system suggests the charging station closest to the reported obfuscated location. The best case is when the suggested charging station coincides with the ones that the system would have suggested if the actual UAV location was disclosed, as the extra distance to be travelled by the UAV because of the application of location obfuscation is $0$~m. When this is not the case, the extra distance the UAV has to travel should be as small as possible. In the following, we investigate the utility of \sol\ in this use case by evaluating the extra distance travelled by the UAV to reach the suggested charging station with different protocol configurations.
We consider a UAV disclosing locations with average distance values of 25~m, 50~m, and 100~m, and a variable number of deployed charging stations, i.e., 2, 8, and 32, so to investigate the impact of a variable density of such charging stations on the utility of our solution.
We summarize the results of our analysis in Tab.~\ref{tab:extra_distance_charging_stations}. 
\begin{table}[htbp]
\centering
    \caption{Extra distance travelled by the UAV in Use Case \#2, with various configurations.}
    \label{tab:extra_distance_charging_stations}
    \begin{tabular}{p{2.3cm}|p{1.5cm}|p{1.5cm}|p{1.5cm}}
                               & \multicolumn{3}{c}{\textbf{No. of Charging Stations}}       \\ \cline{2-4}
                               & 2                 & 8                  & 32                 \\ \hline
\textbf{Avg. Distance {[}m{]}} & \multicolumn{3}{c}{\textbf{Average Extra Distance {[}m{]}}} \\ \hline
25                             & 0.167             & 0.643              & 1.576              \\
50                             & 0.698             & 2.794              & 6.839              \\
100                            & 2.627             & 10.193             & 23.047            
\end{tabular}
\end{table}

\begin{figure}[htbp]
    \begin{center}
    \includegraphics[width=.7\columnwidth]{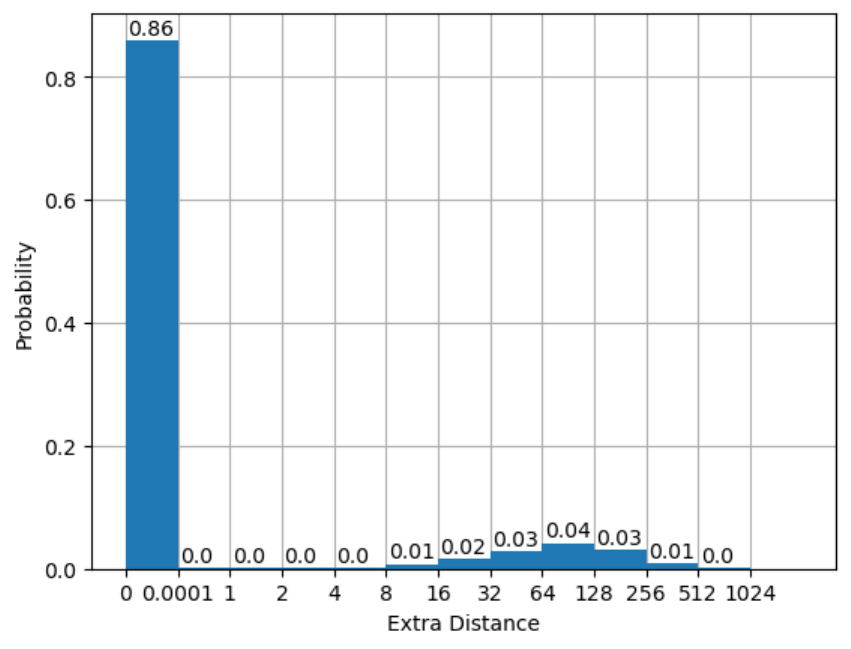}
    \caption{Probability distribution of the average extra distance in Use Case \#2, with an average distance of 100~m and 2 charging stations.}
    \label{fig:histo_charging_station_2_100}
    \end{center}
\end{figure}

\begin{figure}[htbp]
    \begin{center}
    \includegraphics[width=.7\columnwidth]{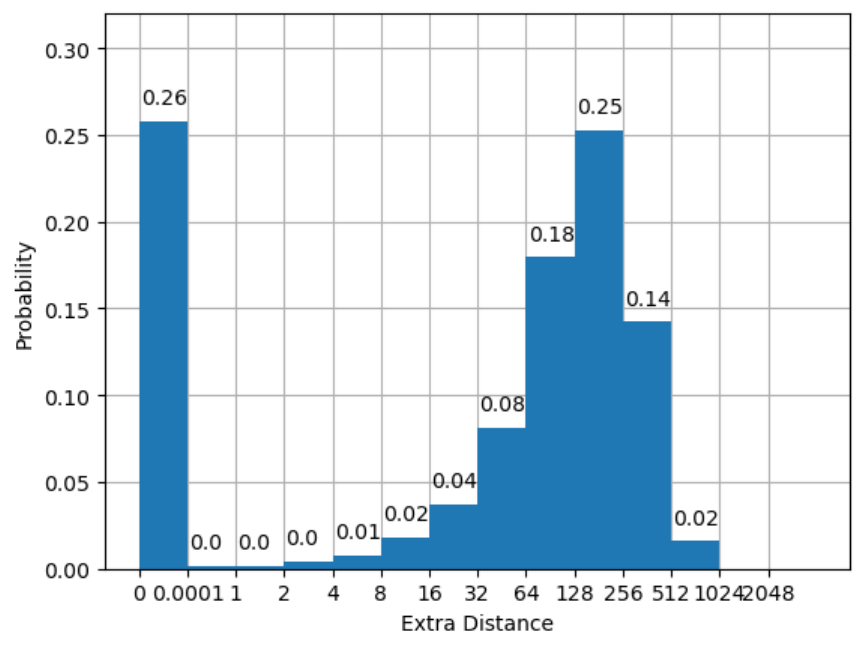}
    \caption{Probability distribution of the average extra distance in Use Case \#2, with average distance of 100~m and 32 charging stations.}
    \label{fig:histo_charging_station_32_100}
    \end{center}
\end{figure}

Increasing the average distance of the locations disclosed by the \ac{UAV} (i.e., increasing \ac{UAV} location privacy) leads to an increase in the average extra distance, i.e., the charging station suggested by the system is not optimal. At the same time, increasing the number of charging stations deployed by the system also increases the average extra distance, as there is more chance that another charging station is closer to the disclosed location than to the actual location.
Asymptotically, if we consider an area completely filled with charging stations, there would always be a charging station coincidental with the UAV's disclosed location. Therefore, the average extra distance asymptotically increases to match the average distance value.
To provide further insights, we show in Figs.~\ref{fig:histo_charging_station_2_100} and \ref{fig:histo_charging_station_32_100} the probability distribution of the extra distance travelled by the UAV considering the average distance of $100$~m with 2 and 32 deployed charging stations, respectively.
When the number of charging stations (density) increases, the probability of picking a suboptimal charging station also increases. We also notice that, excluding the values for the extra distance $0$~m, the remaining values tend to converge to a Gaussian distribution, with mean values coincidental with the average distance. Such a finding confirms our intuition that, asymptotically, the extra distance coincides with the average distance.

\begin{figure}[htbp]
    \begin{center}
    \includegraphics[width=.8\columnwidth]{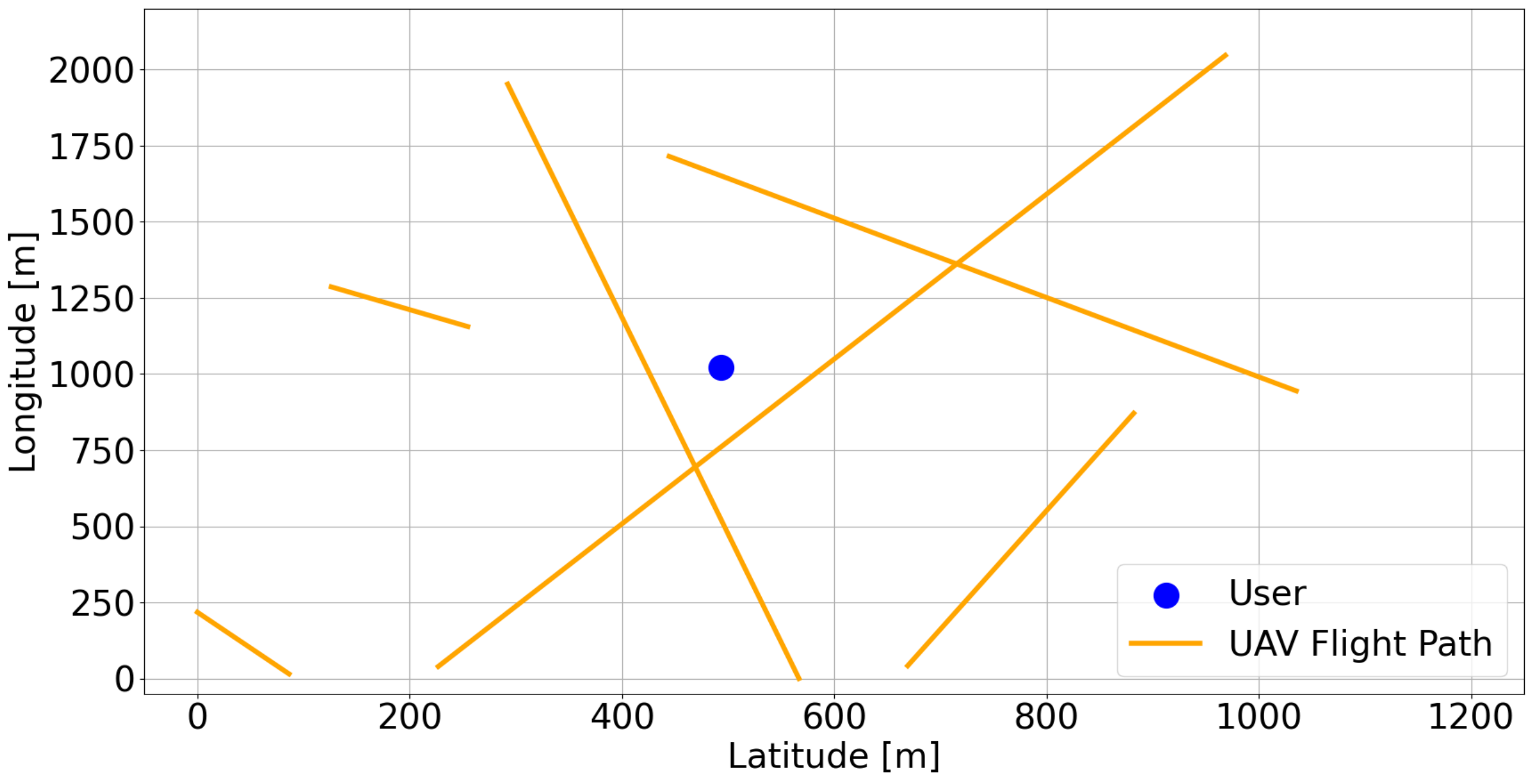}
    \caption{Use Case \#3 Analysis: (6) Flying UAVs providing a service and a user looking for service.}
    \label{fig:exp_service_drones}
    \end{center}
\end{figure}

\begin{figure}[htbp]
    \begin{center}
    \includegraphics[width=.7\columnwidth]{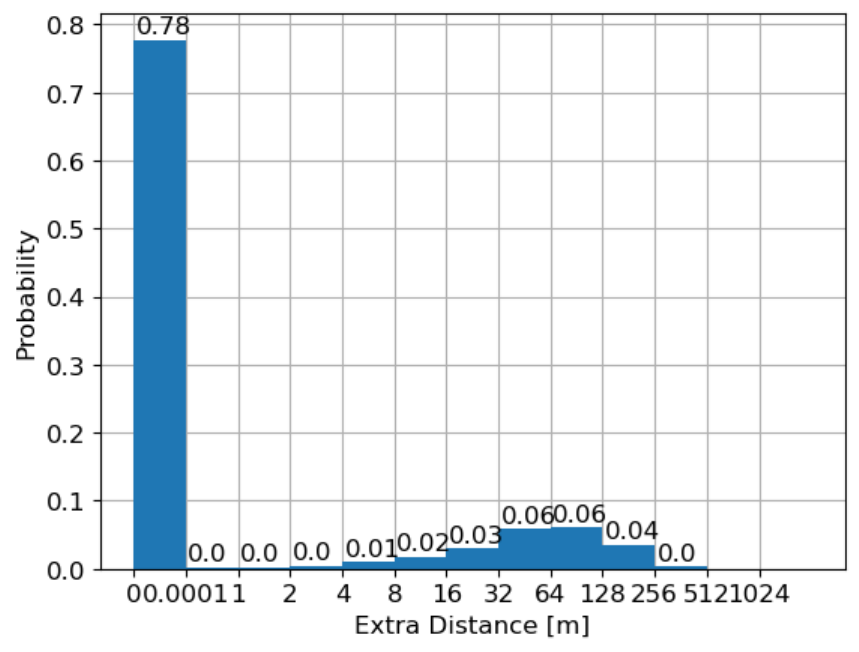}
    \caption{Probability distribution of the average extra distance in Use Case \#3 considering an average distance of 100~m and 2 service UAVs.}
    \label{fig:histo_service_drones_2_100}
    \end{center}
\end{figure}

\begin{figure}[htbp]
    \begin{center}
    \includegraphics[width=.7\columnwidth]{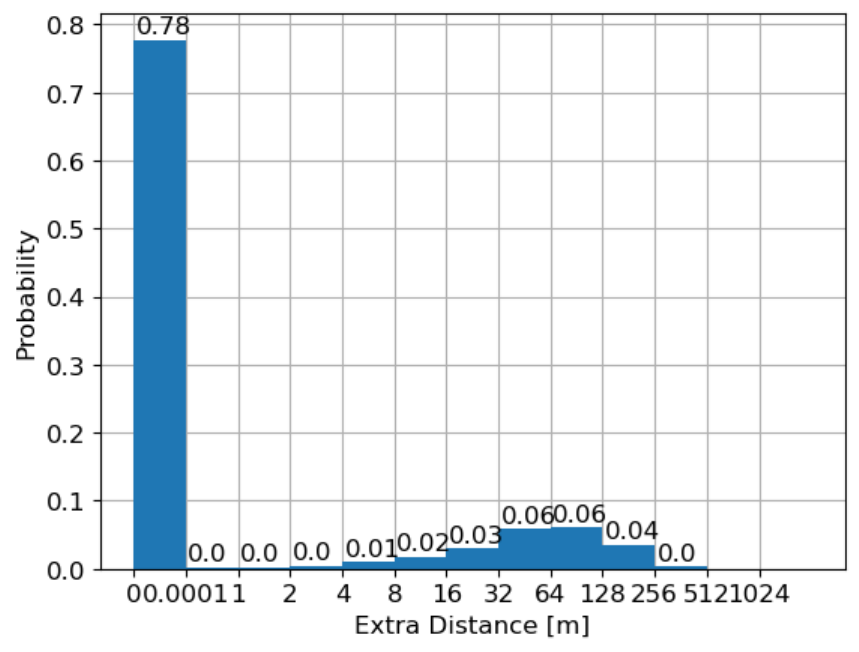}
    \caption{Probability distribution of the average extra distance in Use Case \#3 considering an average distance of 100~m and 10 service UAVs.}
    \label{fig:histo_service_drones_10_100}
    \end{center}
\end{figure}

\indent 
\textbf{Use Case 3: \ac{UAV}-as-a-service.} Consider Use Case \#3 described in Sec.~\ref{subsec:use_cases}, and particularly, the scenario depicted in Fig.~\ref{fig:exp_service_drones}, where the blue marker depicts a user requesting a service and the yellow lines indicate the flight path of various UAVs providing services.
Recall that the user finds the nearest UAV based on the location disclosed by such UAVs through RID messages. Without \sol, the UAV suggested to the user would always be the one closest to the user's location. With \sol\ in place, although being farther than another UAV from the user, a given UAV might disclose a location closer to the user, being selected as the one providing the service to the user. In such cases, the wireless signals emitted by the user to the UAV propagate over an extra distance than the optimal case. In the following, we investigate the utility of \sol\ in this use case by evaluating the extra distance the signal from the user has to travel to reach the suggested UAV based on the reported obfuscated location with different protocol configurations.
We consider a variable number of deployed UAVs, i.e., 2, 6, and 10, and various values of average distance in the location disclosed by such UAVs, i.e., 25~m, 50~m, and 100~m. We summarize the results of our analysis in Tab.~\ref{tab:extra_distance_service_drones}.
\begin{table}[htbp]
    \centering
    \caption{Extra distance travelled by the signal in Use Case \#3, with various configurations.}
    \label{tab:extra_distance_service_drones}
    \begin{tabular}{p{2.3cm}|p{1.5cm}|p{1.5cm}|p{1.5cm}}
                                   & \multicolumn{3}{c}{\textbf{No. of Service UAVs}}       \\ \cline{2-4}
                                   & 2                 & 6                  & 10                 \\ \hline
    \textbf{Avg. Distance {[}m{]}} & \multicolumn{3}{c}{\textbf{Average Extra Distance {[}m{]}}} \\ \hline
    25 & 0.316 & 0.924 & 1.196 \\
    50 & 1.136 & 2.985 & 4.460 \\
    100 & 4.8432 & 11.764 & 16.636           
    \end{tabular}
\end{table}
We notice that the results reported in Tab.~\ref{tab:extra_distance_service_drones} for Use Case \#3 align with the ones in Tab.~\ref{tab:extra_distance_charging_stations} for Use Case \#2. Increasing the average distance, i.e., increasing \acp{UAV}' location privacy, leads to a higher extra distance, i.e., suboptimal performances of the location-based service. Similarly, increasing the number of UAVs increases the average extra distance, leading to suboptimal performances. We can apply the same reasoning as for Use Case \#2 to reason about the asymptotic behavior of the extra distance in Use Case \#3: assuming there is a UAV providing service on top of each possible location (i.e., an infinite number of UAVs), the average extra distance would match precisely the average distance.
\textcolor{black}{We provide more insights into the service degradation caused by privacy preservation by reporting the additional attenuation incurred by a sample signal in an exemplary deployment. We consider a transmitter emitting signals at the frequency of 2.5 GHz, located at random coordinates of latitude 42.3001 and longitude -71.3504. We also considered a random receiver, operating on the same frequency, located at latitude 42.3467 and longitude -71.0972. For such an example deployment, we consider the propagation model provided by Matlab for heavy rain, and we compute the path loss at the receiver site using such a propagation model. Then, to show the impact of our solution, we computed the path loss experienced by the receiver when considering the average extra distance in the worst case of Tab.~\ref{tab:extra_distance_service_drones}, i.e., 100~m of average distance in the locations disclosed by the \ac{UAV} and a number of 10 service UAVs. Without applying our solution, the path loss experienced by the receiver is 127.320793 db. Such a path loss becomes 127.320796 db at an average extra distance of 1.19 6m (average distance of 25 m in the locations disclosed by the \acp{UAV}), 127.320801 db at an average extra distance of 4.46 m (average distance of 50 m in the locations disclosed by the \acp{UAV}), and finally 127.320822 db at an average extra distance of 16.636 m (average distance of 100 m in the locations disclosed by the \acp{UAV}). Therefore, even in most privacy-preserving scenario (100 meters of average distance in the locations disclosed by the \acp{UAV}), the users experience a degradation of the received signal of only 0.000029 db, which is clearly very limited, thus confirming the value of our solution.}
Finally, we notice a similarity between Use Case \#2 and Use Case \# 3 also when analyzing the probability distribution of the extra distance for specific configurations, as shown in Figs.~\ref{fig:histo_service_drones_2_100} and \ref{fig:histo_service_drones_10_100} considering an average distance of $100$~m with 2 and 10 UAVs, respectively.
When the UAVs providing services increase, the user's probability of picking the suboptimal UAV also increases. Excluding the values for the extra distance $0$~m, the remaining values tend to converge to a Gaussian distribution, with mean values coincidental with the value of the average distance on the \ac{UAV}. This finding aligns with our intuition that, asymptotically, the extra distance coincides with the average distance.
\section{Discussion}
\label{sec:discussion}


{\bf Summary.} Through our extensive performance assessment (Sec.~\ref{sec:evalution_and_results}), we experimentally proved that \sol\ allows to generate and deliver RID messages protecting the \acp{UAV}' location privacy within $0.16$~s, i.e., below the threshold of $1$~s set by the \ac{RID} regulation~\cite{remote_id}. The RAM consumption and communication overhead of our solution are also negligible, guaranteeing the potential smooth integration of our solution into running RID-compliant deployments.

{\bf Energy Considerations.} As discussed in Sec.~\ref{subsec:experiment_1_overhead_baseline}, a single RID message requires $0.149$ mW and $0.216$ mW of power for the \acp{EC} BN254 and BLS48556 on the \ac{RPI}, respectively. Therefore, a single RID message requires $0.0140$ mJ and $0.0147$ mJ of energy for \acp{EC} BN254 and BLS48556, respectively. We can express such values in ampere-hours as $Ah=E/(V\cdot 3600)$, where $Ah$ is the ampere-hours (Ah), $E$ is the energy usage (J), and V is the voltage (V) of the battery of the UAV, e.g., $7.7$~V for the DJI Mini 2~\cite{dji_mini_2}. Therefore, a single RID message secured through our solution requires from the UAV $0.000505$~$\mu$Ah and $0.000530$ $\mu$Ah for \acp{EC} BN254 and BLS48556, respectively. As the battery capacity of the DJI Mini 2 is $2250$ mAH, \sol\ consumes between $0.0224\%$ and $0.0236\%$ of the DJI Mini 2 total battery, confirming its limited energy overhead.

{\bf Compliance to Requirements.} Here, we look back to the requirements defined in Sec.\ref{subsec:requirements} to evaluate the compliance of our solution. The requirement \emph{R1, Location Obfuscation} is fulfilled by integrating and extending the solution in~\cite{xiao2015_ccs}.
As per the requirement \emph{R2, Trajectory Privacy}, as shown in Sec.~\ref{sec:security_analysis}, our system is as secure as the \ac{PIM} method, so it offers trajectory privacy. Our solution does not require any Internet connection on the UAV, so it also fulfils the requirement \emph{R3, No Persistent Internet Connection}. Internet connection might be required for some use cases discussed in Sec.~\ref{subsec:use_cases}, but the absence of such a connection does not prevent the \ac{UAV} from running \sol. As per the requirement \emph{R4, Maximum Message Generation Time.}, as shown in Sec.~\ref{subsec:experiment_1_overhead_baseline}, the maximum time our solution takes to generate a RID message is always less than $0.16$ seconds. Finally, as per the requirement \emph{R5, Maximum Communication Overhead}, the modified RID message disclosed through \sol\ is at most 227 bytes, below the MTU of WiFi of $2,304$~~bytes \cite{ieee2012}.

\textbf{RID Extension and Backward Compatibility.} To protect the UAVs' location privacy, \sol\ extends the format of standard RID messages. Instead of delivering the actual UAV location, we deliver an \emph{obfuscated location}, i.e., a location loosely correlated with the actual one, as described in Sec.~\ref{sec:core_proposal}. We also extend the standard RID message with \emph{encrypted location reports}, containing the encryption of the actual UAV location through the public key of the \ac{USS}. If necessary (see use case \#1 in Sec.~\ref{subsec:use_cases}), the \ac{USS} only can decrypt the UAV's actual location. Note that regular observers do not necessarily need to be updated to be able to decode the new cryptographic fields. General Public Observers can neglect those fields and only refer to the disclosed obfuscated location for the valid \ac{UAV} location. The only observers requiring software updates are the Public Safety Observers, likely connected to the USS for potential disclosure of the \acp{UAV}' actual location, if necessary. Therefore, our solution preserves backward compatibility with current deployments as much as possible while also enhancing the location privacy of RID-compliant \acp{UAV}.

\textcolor{black}{\textbf{Privacy and Utility Considerations.}} Our experimental assessment (Sec.~\ref{sec:evalution_and_results}) provides insights into the trade-off between location privacy and utility to \ac{LBS}. 
Consider \acp{UAV} obfuscating their location by an average distance of $100$ meters. For the use case \#1, CI operators using a small \acl{WA} achieve above $0.87$ \ac{TP} rate and only $0.14$ of \ac{FP} rate. For the use case \#2, considering 10 deployed charging stations, the average extra distance travelled by the \ac{UAV} in an area of $1.5$~km $\times 2.6$~km $\times 40$~m is only $23$~m. Finally, for use case \#3, considering the same area and 10 deployed service \acp{UAV}, the average extra distance travelled by the signal is only $16$~m. \textcolor{black}{Note that such utility metrics are domain-specific, since they show the applicability of our solution to the specific use cases.} Overall, such results highlight that achieving higher location privacy than the standard RID regulation is possible while also guaranteeing meaningful utility of the disclosed location to \acp{LBS}. Recall that, whenever needed, relevant aviation authorities can decrypt the location report to recover the precise \ac{UAV} location. \textcolor{black}{We also highlight that, as already demonstrated by several papers in the literature, e.g.,~\cite{andres2013_ccs} and~\cite{mendes2020_pets}, the effective location privacy provided by \emph{any} privacy-preserving mechanism in the presence of frequent broadcasts of location decreases. They also already demonstrate that privacy decreases more when more messsages ($k$) are collected by the adversary. This is a fundamental limitation of the scenario where (location) privacy protection techniques are applied, and not a problem of our solution. To mitigate such a leakage in our scenario, knowing that \ac{RID} messages are delivered once a second and a static adversary could receive them over a given area, the operator of the \ac{UAV} could decrease the value of $\epsilon$ on purpose, so that, in the worst case, the effective privacy protection of the $k$ messages disclosed to the adversary does not exceed a given bound.}

\textcolor{black}{
\textbf{Limitations.} Our solution applies to \acp{RPAS}, i.e., \acp{UAV} with a pilot (either human or \ac{CS}). For such \acp{UAV}, assuming uniform probability of the next movement is realistic, particularly for entertainment applications and video making, where \acp{UAV} could move unpredictably in any direction.
We acknowledge that, especially for autonomous \acp{UAV} on a mission, such an assumption could be less realistic, requiring a change in the values of the cells in the transition matrix. Although such values do not change the overhead of our solution, they could affect the privacy provided to the autonomous \ac{UAV}.
The performance evaluation presented in Sec.~\ref{subsec:experiment_1_overhead_baseline} uses a single NATO dataset, available at~\cite{flight_data}, since it is the only publicly available dataset of \emph{real \ac{UAV} flights} at the time of this writing. To the best of our knowledge, all other \ac{UAV} flight datasets available online include synthetic trajectories, which do not represent actual \ac{UAV} flights. More complex flight paths could change the reported performance and utility figures. We also acknowledge that our paper considers only three reference use cases, and it does not cover all possible use cases where \acp{UAV} could be applied.
}

\section{Conclusion and Future Work}
\label{sec:conclusion_future_work}

In this paper, we proposed \sol, a framework providing privacy-preserving location disclosure to RID-compliant \acp{UAV}. Through the integration of a state-of-the-art solution for correlated location disclosure and its extension for the \ac{UAV} scenario, \sol\ allows \acp{UAV} to broadcast through RID messages obfuscated locations, protecting as much as possible trajectory privacy while dealing with processing, communication, memory, and energy constraints of the \acp{UAV}. \sol\ also extends RID with encrypted location reports, allowing aviation authorities to obtain the actual \ac{UAV} location if necessary. We implemented two proofs-of-concept of our solution on a regular laptop and a \ac{RPI}, and we experimentally show the viability of our solution for medium-end RID-compliant \acp{UAV} ($0.16$~s of RID message generation time and $0.0236\%$ of energy for a DJI Mini 2). We also evaluate the utility of the disclosed location for three use cases involving \acp{UAV} usage, achieving a reasonable trade-off between \acp{UAV} location privacy and the quality of the provided service. Overall, our work demonstrates that we can achieve accountability for \ac{UAV} operations without entirely sacrificing privacy, paving the way for a practical real-world regulatory framework that considers all such requirements simultaneously. 

\section*{Acknowledgements}
\label{sec:ack}

This work has been supported by the INTERSECT project, Grant No. NWA.1162.18.301, funded by the Netherlands Organisation for Scientific Research (NWO). The findings reported herein are solely responsibility of the authors.

\bibliographystyle{IEEEtran}
\bibliography{refs}

@misc{remote_id, 
    author = {FAA},
    title={Remote Identification of Unmanned Aircraft},
    url ={https://www.faa.gov/sites/faa.gov/files/2021-08/RemoteID_Final_Rule.pdf},
    year={2021},
    note={Accessed: Sep-05-2025}
}

@inproceedings{andres2013_ccs,
  title={{Geo-indistinguishability: Differential privacy for location-based systems}},
  author={Andr{\'e}s, Miguel E and Bordenabe, Nicol{\'a}s E and Chatzikokolakis, Konstantinos and Palamidessi, Catuscia},
  booktitle={Proc. of ACM conference on Computer \& communications security},
  pages={901--914},
  year={2013}
}

@inproceedings{chatzikokolakis2014_pets,
  title={{A predictive differentially-private mechanism for mobility traces}},
  author={Chatzikokolakis, Konstantinos and Palamidessi, Catuscia and Stronati, Marco},
  booktitle={14th Int. Symp. on Privacy Enhancing Technologies},
  pages={21--41},
  year={2014},
  organization={Springer}
}

@article{belwafi2022_access,
  title={{Unmanned Aerial Vehicles’ Remote Identification: A Tutorial and Survey}},
  OPTauthor={Belwafi, Kais and Alkadi, Ruba and Alameri, Sultan A and Al Hamadi, Hussam and Shoufan, Abdulhadi},
  author={{K. Belwafi, R. Alkadi, S. Alameri, et al.}},
  journal={IEEE Access},
  volume={10},
  pages={87577--87601},
  year={2022},
  publisher={IEEE}
}

@INPROCEEDINGS{Sampigethaya2009_dasc,
  author={Sampigethaya, Krishna and Poovendran, Radha},
  booktitle={2009 IEEE/AIAA 28th Digital Avionics Systems Conference}, 
  title={{Privacy of future air traffic management broadcasts}}, 
  year={2009},
  volume={},
  number={},
  pages={6.A.1-1-6.A.1-11},
  OPTdoi={10.1109/DASC.2009.5347456}}

@INPROCEEDINGS{svaigen2022_icc,
  author={Svaigen, Alisson Renan and Boukerche, Azzedine and Ruiz, Linnyer B. and Loureiro, Antonio A. F.},
  booktitle={ICC 2022 - IEEE International Conference on Communications}, 
  title={{A Topological Dummy-based Location Privacy Protection Mechanism for the Internet of Drones}}, 
  year={2022},
  volume={},
  number={},
  pages={1-6},
  OPTdoi={10.1109/ICC45855.2022.9838525}}

@ARTICLE{svaigen2022_iotmag,
  OPTauthor={Svaigen, Alisson R. and Boukerche, Azzedine and Ruiz, Linnyer B. and Loureiro, Antonio A.F.},
  author={{A. Svaigen, A. Boukerche, L. Ruiz, et al.}},
  journal={IEEE Internet of Things Mag.}, 
  title={{Design Guidelines of the Internet of Drones Location Privacy Protocols}}, 
  year={2022},
  volume={5},
  number={2},
  pages={175-180},
  OPTdoi={10.1109/IOTM.008.2100131}}

@inproceedings{xiao2015_ccs,
author = {Xiao, Yonghui and Xiong, Li},
title = {{Protecting Locations with Differential Privacy under Temporal Correlations}},
year = {2015},
booktitle = {Proc. of ACM Conf. on Computer and Communications Security},
pages = {1298–1309},
OPTlocation = {Denver, Colorado, USA},
OPTseries = {CCS '15}
}

@article{mendes2020_pets,
author = {Mendes, Ricardo and Cunha, Mariana and Vilela, Joao},
year = {2020},
month = {04},
pages = {379-396},
title = {{Impact of Frequency of Location Reports on the Privacy Level of Geo-indistinguishability}},
volume = {2020},
journal = {Proceedings on Privacy Enhancing Technologies},
doi = {10.2478/popets-2020-0032}
}

@inproceedings{brighente2022_ares,
author = {Brighente, Alessandro and Conti, Mauro and Sciancalepore, Savio},
title = {{Hide and Seek: Privacy-Preserving and FAA-Compliant Drones Location Tracing}},
year = {2022},
booktitle = {Int. Conf. on Availability, Reliability and Security},
OPTisbn = {9781450396707},
OPTpublisher = {Association for Computing Machinery},
OPTaddress = {New York, NY, USA},
OPTurl = {https://doi.org/10.1145/3538969.3543784},
OPTdoi = {10.1145/3538969.3543784},
OPTarticleno = {134},
OPTnumpages = {11},
OPTseries = {ARES '22}
}

@article{xiao2017_vldb,
author = {Xiao, Yonghui and Xiong, Li and Zhang, Si and Cao, Yang},
title = {{LocLok: Location Cloaking with Differential Privacy via Hidden Markov Model}},
year = {2017},
issue_date = {August 2017},
publisher = {VLDB Endowment},
volume = {10},
number = {12},
issn = {2150-8097},
journal = {Proc. VLDB Endow.},
month = {08},
pages = {1901–1904},
numpages = {4}
}

@article{wigchert2025_comnet,
  title={{Detection of Aerial Spoofing Attacks to LEO Satellite Systems via Deep Learning}},
  author={Wigchert, Jos and Sciancalepore, Savio and Oligeri, Gabriele},
  journal={Computer Networks},
  pages={111408},
  year={2025},
  publisher={Elsevier}
}

@article{xiong2019_ijdsn,
  title={Locally differentially private continuous location sharing with randomized response},
  author={Xingxing Xiong and Shubo Liu and Dan Li and Jun Wang and Xiaoguang Niu},
  journal={International Journal of Distributed Sensor Networks},
  year={2019},
  volume={15}
}

@inproceedings{sciancalepore2022_acsac,
  title={{Privacy-Preserving Trajectory Matching on Autonomous Unmanned Aerial Vehicles}},
  author={Sciancalepore, Savio and George, Dominik Roy},
  booktitle={Proceedings of the 38th Annual Computer Security Applications Conference},
  pages={1--12},
  year={2022}
}

@article{jiang2021_csur,
  title={{Location privacy-preserving mechanisms in location-based services: A comprehensive survey}},
  OPTauthor={Jiang, Hongbo and Li, Jie and Zhao, Ping and Zeng, Fanzi and Xiao, Zhu and Iyengar, Arun},
  author={{H. Jiang, J. Li, P. Zhao, et al.}},
  journal={ACM Comput. Surveys},
  volume={54},
  number={1},
  pages={1--36},
  year={2021},
  publisher={ACM New York, NY, USA}
}

@article{zhang2019_tifs,
  title={Online Location Trace Privacy: An Information Theoretic Approach},
  author={Wenjing Zhang and Ming Li and Ravi Tandon and Hui Li},
  journal={IEEE Transactions on Information Forensics and Security},
  year={2019},
  volume={14},
  pages={235-250}
}

@article{hua2017_tifs,
  title={A Geo-Indistinguishable Location Perturbation Mechanism for Location-Based Services Supporting Frequent Queries},
  author={Jingyu Hua and Fengyuan Xu and Sheng Zhong},
  journal={IEEE Transactions on Information Forensics and Security},
  year={2017},
  volume={13},
  pages={1155-1168}
}

@inproceedings{cao2019_icde,
	OPTdoi = {10.1109/icde.2019.00153},
	OPTurl = {https://doi.org/10.1109%2Ficde.2019.00153},  
	year = {2019},
	month = {04},  
	publisher = {{IEEE}},  
	author = {Yang Cao and Yonghui Xiao and Li Xiong and Liquan Bai},  
	title = {{PriSTE}: From Location Privacy to Spatiotemporal Event Privacy},  
	booktitle = {Int. Conf. on Data Engineering ({ICDE})}
}

@misc{aeroscope,
  author = {{Drone XL}},
  title = {{DJI drones transmit location data unencrypted; ‘open source AeroScope’ in the making}},
  year={2022},
  howpublished={ \url{https://dronexl.co/2022/05/09/dji-aeroscope/}}
}

@misc{statista_dronesMarket,
  author = {{Statista}},
  title = {{Consumer drone unit shipments worldwide from 2020 to 2030}},
  year={2020},
  howpublished={ \url{https://www.statista.com/statistics/1234658/worldwide-consumer-drone-unit-shipments/#:~:text=The%20total%20number%20of%20consumer,unit%20shipments%20globally%20by%202030}}
}

@misc{statista_dronesUS,
  author = {{Statista}},
  title = {{Unmanned aircraft systems (UAS)/drones registered in the United States as of July 2023, by category}},
  year={2023},
  howpublished={ \url{https://www.statista.com/statistics/1221517/uas-drone-registrations-united-states/}}
}

@ARTICLE{Shakhatreh2019_access,
  OPTauthor={Shakhatreh, Hazim and Sawalmeh, Ahmad H. and Al-Fuqaha, Ala and Dou, Zuochao and Almaita, Eyad and Khalil, Issa and Othman, Noor Shamsiah and Khreishah, Abdallah and Guizani, Mohsen},
  author={{H. Shakhatreh, A. Sawalmeh, A. Al-Fuqaha, et al.}},
  journal={IEEE Access}, 
  title={{Unmanned Aerial Vehicles (UAVs): A Survey on Civil Applications and Key Research Challenges}}, 
  year={2019},
  volume={7},
  number={},
  pages={48572-48634},
  OPTdoi={10.1109/ACCESS.2019.2909530}}

@misc{raspberry_pi, 
title={Raspberry Pi 3 Model B+}, 
url={https://www.raspberrypi.com/products/raspberry-pi-3-model-b-plus/}, 
journal={Raspberry Pi}, 
note={Accessed: Sep-05-2025}
}

@misc{djiMatrice,
    url={https://enterprise.dji.com/matrice-350-rtk?site=enterprise&from=nav},
    title={DJI Matrice 300 RTK},
    journal={{DJI}},
    note={Accessed: Sep-05-2025}
}

@inproceedings{george2023_sacmat,
  title={{Privacy-Preserving Multi-Party Access Control for Third-Party UAV Services}},
  author={George, Dominik Roy and Sciancalepore, Savio and Zannone, Nicola},
  booktitle={Proceedings of the 28th ACM Symposium on Access Control Models and Technologies},
  pages={19--30},
  year={2023}
}

@misc{drip,
    title={Drone Remote ID Protocol (drip)},
    url={https://datatracker.ietf.org/wg/drip/about/},
    note={Accessed: Sep-05-2025}
}

@inproceedings{tedeschi2021_acsac,
author = {Tedeschi, Pietro and Sciancalepore, Savio and Di Pietro, Roberto},
title = {{ARID: Anonymous Remote IDentification of Unmanned Aerial Vehicles}},
year = {2021},
OPTisbn = {9781450385794},
OPTpublisher = {Association for Computing Machinery},
OPTaddress = {New York, NY, USA},
OPTurl = {https://doi.org/10.1145/3485832.3485834},
OPTdoi = {10.1145/3485832.3485834},
booktitle = {Annual Computer Security Applications Conference},
pages = {207–218},
OPTnumpages = {12},
OPTkeywords = {Remote Identification, Prototyping, Authentication, Anonymity, Unmanned Aerial Vehicles},
OPTlocation = {Virtual Event, USA},
OPTseries = {ACSAC '21}
}

@ARTICLE{wisse2023_iotj,
  author={Wisse, Eva and Tedeschi, Pietro and Sciancalepore, Savio and Pietro, Roberto Di},
  journal={IEEE Internet of Things J.}, 
  title={{A2RID -Anonymous Direct Authentication and Remote Identification of Commercial Drones}}, 
  year={2023},
  volume={},
  number={},
  pages={1-1},
  OPTdoi={10.1109/JIOT.2023.3240477}}

@misc{rfc9153,
    series =    {Request for Comments},
    number =    9153,
    howpublished =  {RFC 9153},
    publisher = {RFC Editor},
    doi =       {10.17487/RFC9153},
    url =       {https://www.rfc-editor.org/info/rfc9153},
        author =    {Stuart W. Card and Adam Wiethuechter and Robert Moskowitz and Andrei Gurtov},
    title =     {{Drone Remote Identification Protocol (DRIP) Requirements and Terminology}},
    pagetotal = 41,
    year =      2022,
    month =     feb,
}

@misc{rfc9434,
    series =    {Request for Comments},
    number =    9434,
    howpublished =  {RFC 9434},
    publisher = {RFC Editor},
    doi =       {10.17487/RFC9434},
    url =       {https://www.rfc-editor.org/info/rfc9434},
        author =    {Stuart W. Card and Adam Wiethuechter and Robert Moskowitz and Shuai Zhao and Andrei Gurtov},
    title =     {{Drone Remote Identification Protocol (DRIP) Architecture}},
    pagetotal = 28,
    year =      2023,
    month =     jul
}

@misc{miracl,
    title={MIRACL Core},
    url={https://github.com/miracl/core},
    note={Accessed: Sep-05-2025}
}

@misc{gcc,
    year = {2023},
    title = {Using the GNU Compiler Collection},
    publisher = {Free Software Foundation},
    URL = {https://gcc.gnu.org/onlinedocs/gcc-13.1.0/gcc/},
    note = {Accessed: Sep-05-2025}
}

@inproceedings{bn254_issues,
  title={Extended tower number field sieve: A new complexity for the medium prime case},
  author={Kim, Taechan and Barbulescu, Razvan},
  booktitle={Annual international cryptology conference},
  pages={543--571},
  year={2016},
  organization={Springer}
}

@article{ecies,
author = {Gayoso Martínez, Víctor and Hernandez Encinas, Luis and Sánchez Ávila, Carmen},
year = {2010},
month = {01},
pages = {7-13},
title = {A Survey of the Elliptic Curve Integrated Encryption Scheme},
volume = {2},
journal = {Journal of Computer Science and Engineering}
}

@InProceedings{ecies_standard,
author={Daniel R. L. Brown},
year={2009},
month={05},
booktitle="Standards for Efficient Cryptography",
title={SEC 1: Elliptic Curve Cryptography},
volume={2.0},
organization={Certicom Corp}
}

@misc{powertop, 
    url={https://www.intel.com/content/www/us/en/developer/articles/tool/powertop-primer.html}, 
    journal={PowerTOP Primer}, 
    author={Pearce, Michael A}, 
    year={2015}, 
    month={Apr},
    note = {Accessed: Sep-05-2025}
}

@article{nist_curves,
  title={Sec 2: Recommended elliptic curve domain parameters},
  author={Qu, Minghua},
  journal={Certicom Res., Tech. Rep. SEC2-Ver-0.6},
  year={1999},
  publisher={Citeseer}
}

@inproceedings{bn254,
  title={{Pairing-Friendly Elliptic Curves of Prime Order}},
  author={Barreto, Paulo SLM and Naehrig, Michael},
  booktitle={International workshop on selected areas in cryptography},
  pages={319--331},
  year={2005},
  organization={Springer}
}

@inproceedings{bls48,
  title={Secure and efficient pairing at 256-bit security level},
  OPTauthor={Kiyomura, Yutaro and Inoue, Akiko and Kawahara, Yuto and Yasuda, Masaya and Takagi, Tsuyoshi and Kobayashi, Tetsutaro},
  author={{Y. Kiyomura, A. Inoue, Y. Kawahara, et al.}},
  booktitle={Int. Conf. on Applied Cryptography and Network Security},
  pages={59--79},
  year={2017},
  organization={Springer}
}

@misc{flight_data,
author = {NATO},
year = {2021},
title={Drone identification and tracking},
URL = {https://www.kaggle.com/c/icmcis-drone-tracking/overview},
note = {Accessed: Sep-05-2025}
}

@misc{ieee2012,
  title={{IEEE Computer Society:“Part 11: Wireless LAN Medium Access Control (MAC) and Physical Layer (PHY) Specifications”, IEEE Std 802.11™-2012, The Institute of Electrical and Electronics Engineers}},
  author={IEEE Standards},
  year={2012},
  OPTpublisher={Inc}
}

@misc{dji_mini_2,
author = {DJI},
year = {2023},
title={DJI MINI 2 Specs},
URL = {https://www.dji.com/nl/mini-2/specs},
note = {Accessed: Sep-05-2025}
}

@article{tedeschi2022_tdsc,
  title={{PPCA-Privacy-Preserving Collision Avoidance for Autonomous Unmanned Aerial Vehicles}},
  author={Tedeschi, Pietro and Sciancalepore, Savio and Di Pietro, Roberto},
  journal={IEEE Transactions on Dependable and Secure Computing},
  volume={20},
  number={2},
  pages={1541--1558},
  year={2022},
  publisher={IEEE}
}

@ARTICLE{enayati2023_iotj,
  OPTauthor={Enayati, Saeede and Goeckel, Dennis and Houmansadr, Amir and Pishro-Nik, Hossein},
  author={{S. Enayati, D. Goeckel, A. Houmansadr, et al.}},
  journal={IEEE Internet of Things J.}, 
  title={{Location Privacy Protection for UAVs in Package Delivery and IoT Data Collection}}, 
  year={2023},
  volume={10},
  number={23},
  pages={20586-20601},
  OPTdoi={10.1109/JIOT.2023.3293755}}

@inproceedings{tedeschi2023_acsac,
  title={{Lightweight Privacy-Preserving Proximity Discovery for Remotely-Controlled Drones}},
  author={Tedeschi, Pietro and Sciancalepore, Savio and Di Pietro, Roberto},
  booktitle={Proc. of Annual Computer Security Applications Conference},
  pages={178--189},
  year={2023}
}

@inproceedings{tedeschi2021_spaccs,
  title={{Modelling a Communication Channel under Jamming: Experimental Model and Applications}},
  author={Tedeschi, Pietro and Sciancalepore, Savio and Di Pietro, Roberto},
  booktitle={IEEE Intl Conf on Parallel \& Distributed Processing with Applications, Big Data \& Cloud Computing, Sustainable Computing \& Communications, Social Computing \& Networking},
  pages={1562--1573},
  year={2021},
  OPTorganization={IEEE}
}

@misc{code,
  title = {{Open Source Code of OLO-RID}},
  author={Brighente, Alessandro and Conti, Mauro and Schotsman, Matthijs and Sciancalepore, Savio},
  howpublished = {\url{https://github.com/MSchotsman/OLO-RID}},
  note = {Accessed: Sep-05-2025}
}

@article{sciancalepore2023_iotj,
  title={{Jamming Detection in Low-BER Mobile Indoor Scenarios via Deep Learning}},
  author={Sciancalepore, Savio and Kusters, Fabrice and Abdelhadi, Nada Khaled and Oligeri, Gabriele},
  journal={IEEE Internet of Things Journal},
  year={2023},
  publisher={IEEE}
}


\end{document}